\documentclass[11pt,a4paper]{article}
\pdfoutput=1

\usepackage{jheppub} 

\usepackage[T1]{fontenc}
\usepackage{amssymb,amsmath}
\usepackage{bm}

\usepackage{footnote}

\usepackage{bm}

\usepackage{hyperref}

\usepackage{amssymb,amsmath}

\newcommand{\ud}{\,\mathrm{d}}

\usepackage{xcolor}

\newcommand{\beq}{\begin{equation}}
\newcommand{\beqn}{\begin{eqnarray}}
\newcommand{\eeq}{\end{equation}}
\newcommand{\eeqn}{\end{eqnarray}}

\usepackage{physics}

\usepackage{multirow}

\usepackage{verbatim}

\begin{document}

\title{\center \bf \Huge Scaling attractors \\ in multi-field inflation
}

\author[a]{Perseas Christodoulidis,}
\author[a]{Diederik Roest,}
\author[b,c]{Evangelos I. Sfakianakis}
\affiliation[a]{Van Swinderen Institute for Particle Physics and Gravity, 
University of Groningen, Nijenborgh 4, 9747 AG Groningen, The Netherlands}
\affiliation[b]{Nikhef, Science Park 105, 1098 XG Amsterdam, The Netherlands}
\affiliation[c]{Lorentz Institute for Theoretical Physics, Leiden University, 2333 CA Leiden, The Netherlands}
\emailAdd{p.christodoulidis@rug.nl}
\emailAdd{d.roest@rug.nl}
\emailAdd{e.sfakianakis@nikhef.nl}

\abstract{
Multi-field inflation with a curved scalar geometry has been found to support background trajectories that violate the slow-roll, slow-turn conditions {and thus have the potential to evade the swampland constraints}. In order to understand how generic this  novel behaviour is and what conditions lead to it, we perform a classification of  dynamical attractors of two-field inflation that are of the scaling type. Scaling solutions form a one-parameter generalization of De Sitter solutions with a constant value of the first Hubble flow parameter $\epsilon$ and, as we argue and demonstrate, form a natural starting point for the study of non-slow-roll slow-turn behaviour. 

All scaling solutions can be classified as critical points of a specific dynamical system.  
We recover known multi-field inflationary attractors as approximate scaling solutions and classify their stability using dynamical system techniques. 
In particular, we discover that dynamical bifurcations play an integral role in the transition between geodesic and non-geodesic motion and discuss the ability of scaling solutions to describe realistic multi-field models. We revisit the criteria for background stability and show cases where the usual criteria found in the literature do not capture the background evolution of the system.
 }


\begin{flushright}
	Nikhef  2019-007
\end{flushright}

\maketitle

\section{Introduction} 

Inflation is the leading paradigm for the early universe, providing an elegant entrance to the hot Big Bang as well as a way to seed structure at multiple scales, from galaxy clusters to the CMB. Observations, such as by the {\it Planck} satellite \cite{Akrami:2018odb}, are compatible with single-field models of inflation where the scalar field undergoes slow-roll evolution, but also with a plethora of other models. It is  thus both interesting and necessary  to examine possible inflationary scenarios that go beyond the single-field slow-roll model. 

High-energy theories suggest the existence of multiple scalar fields at energy scales relevant for inflation. Furthermore, the field-space metric in many such theories is not flat: the scalar field kinetic terms ${\cal G}_{IJ}(\phi) \partial \phi^I \partial \phi^J$ are not canonical and the fields in general cannot be re-defined to make the kinetic terms canonical. Moreover, it has recently emerged in a number of investigations \cite{Brown:2017osf,Mizuno:2017idt,Achucarro:2015rfa, Christodoulidis:2018qdw, Linde:2018hmx,  Garcia-Saenz:2018ifx, Achucarro:2019pux,
Bjorkmo:2019aev, Bjorkmo:2019fls}, that such curved geometries can support multi-field inflation that differs from the slow-roll, slow-turn paradigm (which moreover can evade the putative swampland conditions \cite{Vafa}, as discussed in Ref.~\cite{Palma}). A proper understanding of these aspects is necessary to correctly interpret the constraints on inflationary models arising from observational data. 

A successful model of inflation should provide predictions that are compatible with   constraints on observables. These are correlation functions of quantum fluctuations evaluated on a classical background trajectory,
 that is obtained as a solution to a system of Klein-Gordon equations for the scalar fields\footnote{There has been a growing interesting for inflationary models involving gauge fields (see e.g. Refs.~\cite{Adshead:2016omu, Adshead:2012kp, Maleknejad:2011jw, Namba:2013kia, Adshead:2017hnc});  however, in such cases, a scalar degree of freedom that takes the role of the inflaton can be identified.}. The first step of inflationary model building is therefore to find solutions of the scalar field equations on a self-consistently expanding FLRW metric. As this system is non-linear, there is no general prescription to tackle the problem\footnote{There are a few examples in the literature which admit general analytical solutions \cite{Salopek:1990jq,Chimento:1998ju,Bertacca:2007ux,Piedipalumbo:2011bj,Basilakos:2011rx,Paliathanasis:2014zxa,Zhang:2009mm,Paliathanasis:2014yfa,Anguelova:2018vyr,Christodoulidis:2018msl,Paliathanasis:2018vru}. These are completely integrable systems where the mini-super Lagrangian has Noether symmetries, and have been classified in Refs.~\cite{Tsamparlis:2011wg,Tsamparlis:2011wf,Tsamparlis:2011nsv}.}; furthermore, general solutions could introduce initial condition dependence on the inflationary predictions, making it generically very hard to disentangle dynamics (e.g.~initial conditions) from theories (e.g.~scalar potentials). 

A feature of inflation that enhances its  predictive power  is the existence of dynamical attractors: for a given model, a large number of solutions will converge (come exponentially close)  to a particular solution that lives on a hypersurface of lower dimensionality than the full problem. The existence of these dynamical attractors is closely related to the expansion of space-time and the induced Hubble friction. In addition, one can often employ simple approximations to derive the (lowest-order) predictions of such attractors. Chief amongst these approximation schemes is the slow-roll 
for a single field, which neglects the second time derivative of the scalar field. In multi-field scenarios this can be generalized to the slow-roll, slow-turn 
neglecting the covariant acceleration of the scalar fields. As mentioned before, this approximation has been found to be violated in multi-field scenarios with specific scalar geometries: the dynamical evolution approaches a different inflationary trajectory. For this reason it would be valuable to have a broader understanding of dynamical attractors in such models. 

In this paper we focus on a more general set of inflationary solutions that we refer to as scaling attractors, continuing the investigation of Ref.~\cite{Christodoulidis:2018msl} where the analysis was limited to flat field-space manifolds. The definition of scaling solutions is that they have a constant parameter $\epsilon = \ud \log(H) / \ud N$, defined in terms of the Hubble parameter $H$ and the number of e-folds $N$. Higher-order Hubble flow parameters $\epsilon_{i+1} = \ud \log(|\epsilon_i|) / \ud N$ \cite{Terrero} are zero in this approximation. Note that this generalizes the slow-roll notion, which can be seen as a deformation of De Sitter space-time with constant $H$ and hence vanishing $\epsilon$. Scaling solutions are therefore a one-parameter generalization of the maximally symmetric De Sitter space-time, with a power-law expansion of space-time and a constant ratio between the kinetic and potential energy of the scalar fields \cite{Lucchin:1984yf}.

Scaling solutions draw their interest from a number of perspectives. For one, they generalize the usual (non-exact) slow-roll slow-turn approximation, which can be seen as small deviations from scaling solutions \cite{Christodoulidis:2018msl}. Secondly, as we will demonstrate, they are exact solutions that can be  conveniently formulated as critical points of dynamical systems. Therefore they can be attractors (in the mathematical sense) that describe the late-time behaviour of large classes of inflationary solutions. Moreover, in the multi-field models with curved geometry that display behaviour away from slow-roll slow-turn \cite{Brown:2017osf, Mizuno:2017idt, Achucarro:2015rfa, Christodoulidis:2018qdw, Linde:2018hmx, Garcia-Saenz:2018ifx, Achucarro:2019pux,Bjorkmo:2019aev}, their inflationary phase can be approximated by such a scaling attractor. A full classification of scaling solutions therefore provides a list of the different possible background behaviours, to which the various inflationary scenarios\footnote{A complementary, more physical discussion of this connection was put forward in Ref.~\cite{Christodoulidis:2019mkj}. The present work can be seen as an explicit illustration that the full catalogue of scaling solutions conforms to this description.} asymptote during a large number of e-folds (before the end of inflation sets in and reheating takes over). 

While the above discussion is limited to the background behaviour, inflation draws its power from predicting correlations of quantum fluctuations on these backgrounds. In a number of the inflationary scenarios mentioned above, such an analysis has already been performed and has been phrased in terms of an effective field theory (EFT)  of the fluctuations during inflation. Despite scaling solutions having $\ud \epsilon / \ud N =0$ by construction, studying deformations away from them can still give useful results for the evolution of fluctuations.

This paper is organized as follows. Sec.~\ref{sec:singlefield} provides an opening discussion of the classification of scaling solutions for a single field, including the formulation in terms of a dynamical system and the relation to actual inflationary scenarios. Special attention is given to the ability of scaling expressions to describe more general behaviour and their relation to the corresponding slow-roll analysis.
Sec.~\ref{sec:multifield_def} outlines the similar notions for a general multi-field model, with particular emphasis on the role of the scalar (field-space) geometry. Throughout most of the subsequent sections, we focus on field-space manifolds with one transitively acting isometry and provide a classification of scaling attractors. Most exact scaling solutions are constructed for potentials 
that exhibit exponential dependence on the fields.
In Sec.~\ref{sec:3d} we study systems with one integral of motion, which can be though of as generalizations of systems with a rotationally symmetric potential. Connections to existing inflationary models \cite{Brown:2017osf, Mizuno:2017idt} are explained.
Sec.~\ref{sec:4d} addresses the more general case of systems with arbitrary potential gradients, also relaxing the requirement for an exponential dependence for one of the two fields. Analogies to recently discovered inflationary trajectories \cite{Achucarro:2015rfa,Christodoulidis:2018qdw,Linde:2018hmx, Garcia-Saenz:2018ifx, Achucarro:2019pux} are drawn and the relation between scaling and slow-roll fast-turn solutions is explained. We relax our assumption for the existence of a field-space isometry and discuss the resulting dynamics in Sec.~\ref{sec:noniso}.
We offer our conclusions and outlook in Sec.~\ref{sec:summary}.
  
 \medskip
 
{\bf Note added:} Upon completion of this manuscript, the preprint \cite{Cicoli:2019ulk} appeared which has some overlap with our discussion  in Sec.~\ref{sec:3d}, albeit in a somewhat different context.
 
\section{Single-field scaling attractors}
\label{sec:singlefield}

\subsection{Critical points analysis}

We  start with the well-known case of inflation driven by a single scalar field. Instead of the usual second-order Klein-Gordon equation, we  consider the equivalent dynamical system formulation consisting of two first-order equations, written in terms of the field $\phi$ and its normalized velocity $v \equiv \phi' = d \phi / dN$, defined as the velocity with respect to the $e$-folding number $N$. The velocity $v$ is related to the field derivative with respect to   cosmic time  via $\phi' =\dot{\phi} / H$. The Hubble parameter can be expressed in terms of these variables, $H = H(\phi,v)$, for any positive potential $V$ through the relation $V = (3 - \epsilon) H^2$. The latter involves the  first slow-roll parameter $\epsilon$ which is related to the the normalized velocity via\footnote{
All field values are in units of the reduced Planck mass, or equivalently we set $M_{\rm Pl}=1$.}

\begin{equation}
\epsilon = \frac{v^2}{2} =  {3 K\over K+V} \,,
\end{equation}
where $K$ and $V$ denote the kinetic and potential energy respectively; note that positive potentials lead to $0\le \epsilon \le 3$ and thus $0\le v\le \sqrt{6}$.
Scaling solutions have constant $\epsilon$ by definition and thus have a constant velocity in terms of $e$-folds. 

The flow of this dynamical system is defined by
\begin{subequations}
	\label{eq:2Dsystem}
	\begin{eqnarray}
	\label{eq:2Dsystem1}
	\phi' &=& v \,, \\
	\label{eq:2Dsystem2}
	\quad v'& =& - (3 - \tfrac12 v^2) \left (v + p \right) \,,
	\end{eqnarray}
\end{subequations}
where the scalar potential  provides a driving force via
\beq
p \equiv {\ud (\ln V) \over \ud \phi} \,, 
\eeq
which is related to the potential first slow-roll parameter $\epsilon_V = \tfrac12 p^2$. An important special case is that of an exponential potential, with $p$ constant, for which the velocity subspace of Eq.~\eqref{eq:2Dsystem2} decouples from Eq.~\eqref{eq:2Dsystem1} and can be studied separately. We therefore first focus on this particularly clear case.

For a constant velocity, the right-hand-side of Eq.~\eqref{eq:2Dsystem2} has to vanish, leading to three distinct critical points\footnote{We use the term ``critical point'' to denote the point at which the derivative of a quantity vanishes. For the dynamical system a critical point  satisfies ${\boldsymbol{x}}'(\boldsymbol{x}_{\rm cp})=0$ whereas for the potential $V' (\phi_{\rm cp} )=0$.} for different values of $p$ (assuming $V = \Lambda \exp(p \phi)$):
\begin{itemize}
	\item
	When $p=0$, i.e.~in the case of a constant potential, one can solve   equations \eqref{eq:2Dsystem} by setting $v=0$. This case of zero velocity and zero acceleration corresponds to an exact De Sitter vacuum with a cosmological constant $\Lambda$, undergoing a never-ending exponential expansion $a(t)= a_0 e^{Ht}$, where $H^2 = \Lambda / 3$. 

This solution will be the late-time attractor for all initial conditions: due to the exponential increase of the scale-factor $a$, any initial velocity is quickly redshifted and there is a fast approach to the above attractor. In addition, since the constant potential has an exact shift symmetry, the cyclic coordinate $\phi$ leads to a conservation law for the associated conjugate momentum $\partial {\cal L} / \partial \dot\phi = a^3 \dot \phi$. 
	\item
	More generally, the flow equations \eqref{eq:2Dsystem} have a critical point  solution for $v = -p$, which can be integrated to yield $\phi =\phi_0 - p N$. This solution has a constant velocity resulting from a tug of war between the exponential gradient of the potential and the Hubble friction term.
	Again, a linear stability analysis indicates that this solution is an attractor for $p < \sqrt{6}$ (with the corresponding eigenvalue equal to $\lambda=-3 + p^2/2$).
	\item
	Finally, a general solution for all $p$ is $\epsilon= 3$ and hence $v = \pm \sqrt{6}$. It corresponds to a configuration whose kinetic term dominates the potential energy (referred to as ``kination'' in Ref.~\cite{JudgmentDay}). A linear stability analysis shows that the solution $v = - \sqrt{6}$ will be stable for $p > \sqrt{6}$.  
\end{itemize}
Thus the classification of critical points nicely illustrates the generalization from De Sitter to scaling solutions with $v = -p$. Stability of the latter only distinguishes between $p \lessgtr \sqrt{6}$ and since the problem is one-dimensional, global stability is fully determined by the linearized results. 

The use of these critical points follows from the asymptotical behaviour of this dynamical system. Even though the full solution for arbitrary initial conditions may be impossible to find, the asymptotic one will have a parametric relation between the fields and velocities provided by the critical point. These relations can be used to calculate  inflationary observables in the regime where the scaling solution provides a good approximation for the full inflationary trajectory\footnote{Note that the scaling solution provides an exact solution for a single field with an exponential potential, while the usual slow-role approach is an approximation where one neglects the acceleration term, which leads to
\begin{equation}
  \dot \phi = -{2\over \sqrt 3} {p\over 2} \,e^{{p\over 2} \phi} \, , \qquad \epsilon = {p^2 \over 2 + p^2 / 3} \,.
\end{equation}
While the two approaches match very well for small values of the potential steepness, the slow-roll approximation breaks down for ${\cal O}(1)$ values of $p$, as expected.}.

More general scalar potentials will not feature asymptotic regimes where   exact scaling solutions apply. However, provided the potential supports slow-roll inflation, one can view the dynamics as a series of successive scaling solutions. The evolution of the first slow-roll parameter is   
\begin{equation} \label{doteps}
\dot \epsilon = {\ddot \phi \dot \phi \over H^2} - 2 \epsilon^2H   \,,
\end{equation}
and hence the deviation from scaling is slow-roll suppressed (with small values for the acceleration and slow-roll parameter). The smallness of the slow-roll parameters leads to a slowly varying field-dependent gradient $p$ of a potential:
\begin{equation}
p_{,\phi} = {V_{,\phi \phi} \over V} - \left(  {V_{,\phi} \over V} \right)^2 = \eta_V - 2 \epsilon_V \,.
\end{equation}
Therefore the slow-roll approximation can be considered as an approximate scaling solution with small $p$ and a slowly moving critical point \cite{Christodoulidis:2018msl}. 

\subsection{Comparison to inflationary models}
\label{singlefieldscaling}

In order to illustrate the critical point catalogue, we now  turn to a more realistic scalar potential that can support inflation for some period and then allows for a graceful exit towards a Minkowski minimum. Consider for instance the Starobinsky model
\beq
V(\phi ) = \Lambda \left (1-e^{\alpha \phi} \right )^2
\label{eq:starobinskyV} \,,
\eeq
generalized with an arbitrary parameter $\alpha$. It has a flat plateau in the far left and an exponential behaviour in the far right.

\begin{figure}
	\centering
	\includegraphics[width=0.6\textwidth]{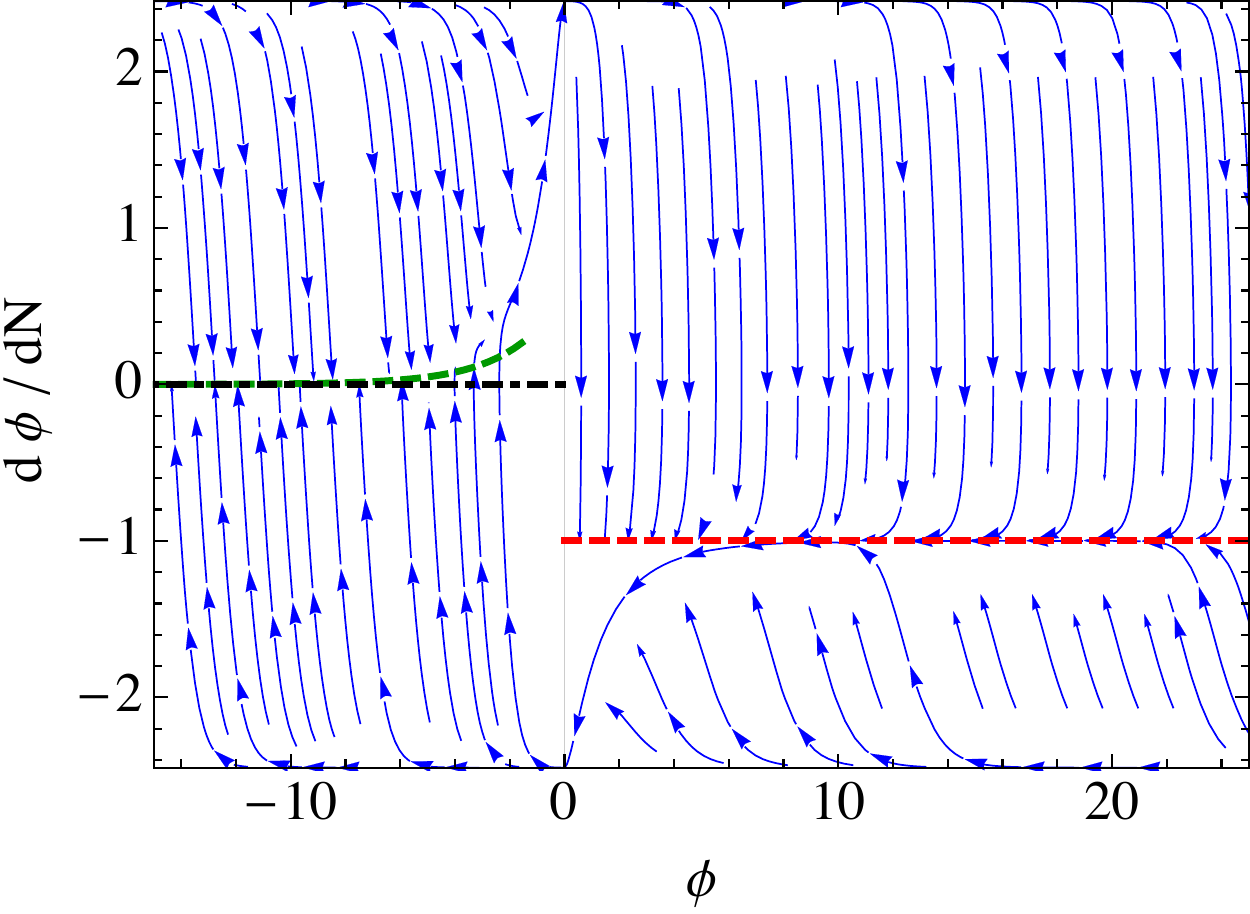}
	\caption{
		The flow of the dynamical system corresponding to a single field with the Starobinsky potential of Eq.~\eqref{eq:starobinskyV} for $\alpha=0.5$. The red dashed line corresponds to the scaling solution $\phi' = -p =-2\alpha$, while the black dot-dashed line corresponds to the De Sitter solution $\phi' = 0$. For illustration purposes we have also included the green dashed line depicting the usual slow-roll solution for evolution on the plateau of the Starobinsky model.}
	\label{fig:StarobinskyFLOW}
\end{figure}

The potential of Eq.~\eqref{eq:starobinskyV} exhibits a flat plateau for large negative values of $\phi$, hence approximating a constant value $V(\phi) \approx \Lambda$ for $\phi \ll -1/\alpha$. The time translation symmetry of an exact De Sitter background in this case  is broken by the rolling of the inflaton field in its potential and can be characterized by the speed of the inflaton, or equivalently by the first slow-roll parameter $\epsilon \equiv \dot H / H^2$. This is illustrated on the left part of the flow diagram of Fig.~\ref{fig:StarobinskyFLOW}, where the horizontal line corresponds to the De Sitter scaling solution and can be seen to be an excellent approximation at sufficiently large field values. However, at smaller field values this scaling approximation breaks down, and one enters a slow-roll, or field-dependent $p(\phi)$ regime.  As the number of e-folds in this regime exceeds that of the observable CMB window, the original scaling phase is not relevant for a study of fluctuations\footnote{Moreover, the semi-classical approximation ignores quantum fluctuations, that can even counter the background motion of the field and lead to a phase of eternal inflation \cite{Linde:1986fc, Vilenkin:1999pi, Guth:1999rh}; see Refs.~\cite{Grain:2017dqa,Pinol:2018euk, Grocholski:2019mot} for recent discussions.}.

Turning to positive field values, the Starobinski potential can be approximated by an exponential $V\propto \exp(2\alpha \phi)$ for $\alpha \phi \gg 1$, and thus can be approximated by a scaling solution of the second type with $v = -p = -2 \alpha$. This is illustrated by the horizontal line on the right hand side of Fig.~\ref{fig:StarobinskyFLOW}. For any $0<p<\sqrt{6}$ this provides a dynamical attractor for a large variety of different initial conditions. In this case the number of e-folds after the actual trajectory has diverged from the scaling solution is far smaller, and hence the fluctution analysis can be based on the scaling attractor. Finally, for $p>\sqrt{6}$ this horizontal line moves downwards and disappears under the limiting case $\epsilon=3$, which indicates that for this parameter range the kinetic solution with $v = - \sqrt{6}$ is the dynamically preferred trajectory.

One can also consider potentials with a different asymptotic field dependence. We first consider a single field model  with a quadratic potential $V(\phi) \propto \phi^2$.  This can be approximated locally around a field value $\phi_*$ by an exponential potential of the form $V(\phi) = V_0 e^{p\phi}$ with $p=2/\phi_*$ and $V_0= \phi_*^2/e^2$. By using the field-dependent value of the potential steepness $p$ and the result $\epsilon = p^2/2$ that was derived for scaling solutions with a constant $p$, we can see that the resulting behavior $\epsilon(\phi)$ agrees very well with the numerically computed slow-roll parameters $\epsilon$ and $\eta$ for a quadratic potential.

\begin{figure}
	\centering
	\includegraphics[width=0.45\textwidth]{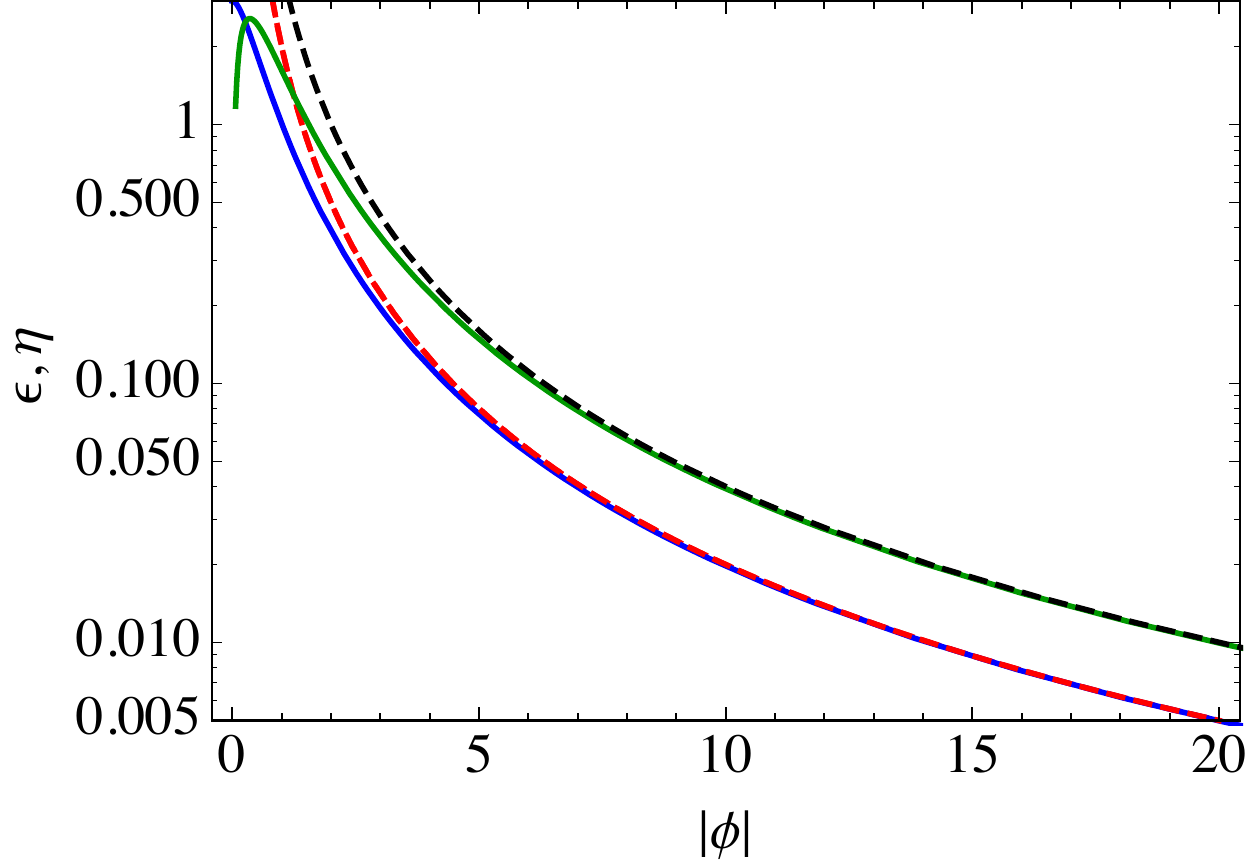}	\includegraphics[width=0.45\textwidth]{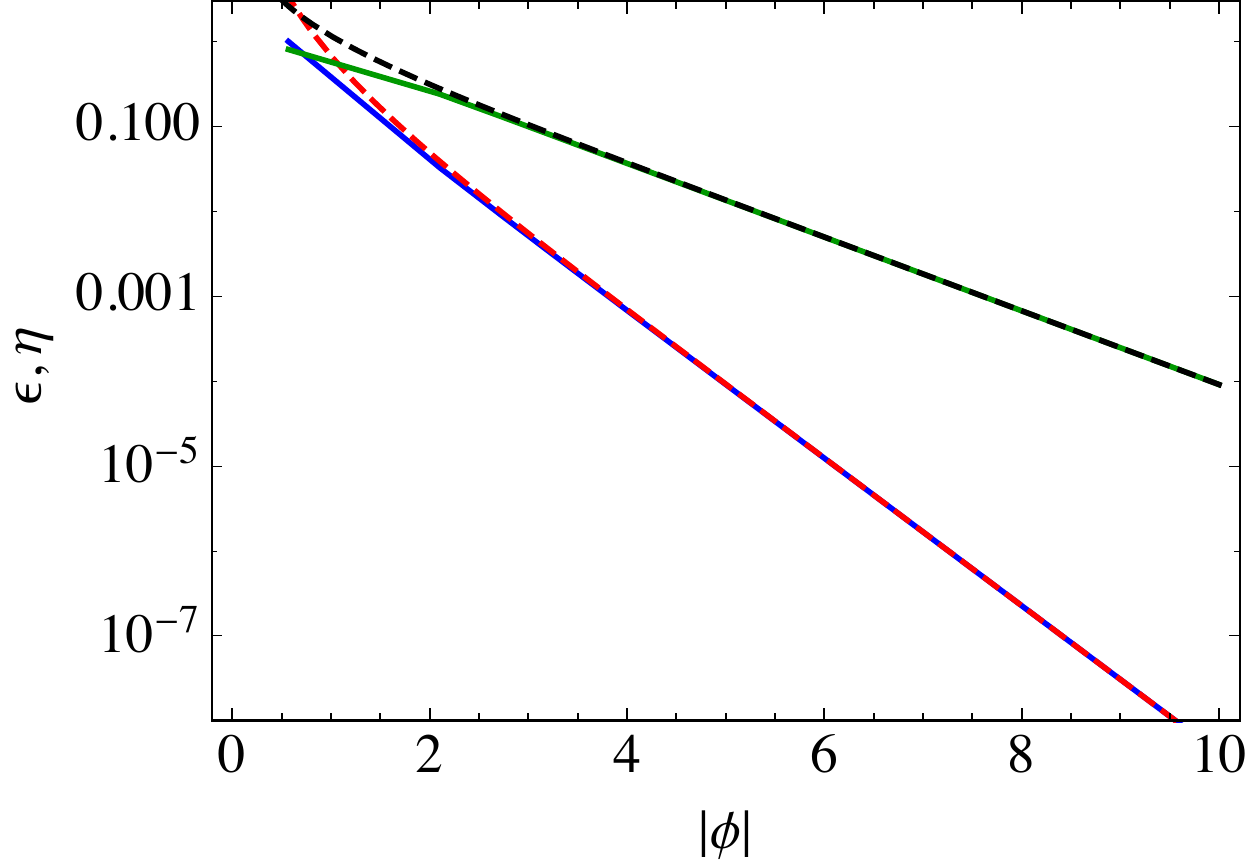}
	\caption{The slow-roll parameters $\epsilon$ (blue \& red) and $\eta$ (green \& black) as a function of the norm of the field value $|\phi|$ computed numerically (solid) and using the scaling relation $\epsilon =p(\phi)^2/2$ (dashed) for a quadratic potential (left panel) and the Starobinsky potential (right panel).}
	\label{fig:quadraticscaling}
\end{figure}

Finally, we would like to illustrate that scaling solutions can also accurately describe a  realistic inflationary background evolution with a concave potential, which is in general preferred by {\it Planck} data. Fig.~\ref{fig:quadraticscaling} shows how, when locally fitting  the plateau of the Starobinsky models with expontential potentials, the resulting (field-dependent) slow-roll parameters $\epsilon$ and $\eta$ from the scaling solution describe the background evolution of the system well in the region of $\eta \ll 1$.

\section{Multi-field inflation}
\label{sec:multifield_def}

\subsection{Dynamical system} \label{subsec:dr}

After showing the applicability of scaling solutions in single-field inflation, we now turn to the main focus of this paper, addressing the analogous questions in multi-field inflation. We consider a system consisting of multiple scalar fields minimally coupled to gravity with Lagrangian density 
\begin{equation}
{\cal L} = \sqrt{-g} \left(\tfrac12  R - \tfrac12 {\cal G}_{IJ}\partial_{\mu} \phi^I\partial^{\mu} \phi^J - V \right) \, ,
\end{equation}
where 
$g_{\mu\nu}$
is the spacetime metric, 
${\cal G}_{IJ}$
is associated with the internal field-space manifold\footnote{A non-trivial field space manifold can also arise if one postulates non-minimal couplings of multiple scalar fields to the Ricci scalar in the Jordan frame and transforms the action to the Einstein frame, where the gravitational sector has the usual Einstein-Hilbert form \cite{Kaiser:2010ps, Kaiser:2010yu}, even if the fields had canonical kinetic terms in the Jordan frame. Models in this class exhibit interesting phenomenology during and after inflation, e.g.~\cite{Kaiser:2013sna, 
DeCross:2016fdz, Sfakianakis:2018lzf, Ema:2016dny}.}  (we use mostly plus signature for the space-time metric, while the field-space metric is positive definite). In an FLRW space-time the background equations of motion are 
\begin{subequations}
	\begin{eqnarray}  \label{kg}
&& {\cal D}_t \dot \phi^I + 3H\dot \phi^I + {\cal G}^{IK}V_{,K}=0 \, ,\\ && \dot H  = - \tfrac12  {\cal G}_{IK} \dot \phi^I \dot \phi^J,
\end{eqnarray}
\end{subequations}
where the covariant directional derivative   and the  Friedmann constraint are given by
\begin{equation}
{\cal D}_t A^I  = \dot A^I + \Gamma^I_{JK}A^J \dot \phi^K \,, \qquad \label{frc}
3H^2  =  \tfrac12 {\cal G}_{IK} \dot \phi^I \dot \phi^J + V.
\end{equation}
The dynamics on a curved manifold is more complex because terms involving field-space Christoffel symbols can act as velocity-dependent forces. It is the interplay between these forces and potential gradient terms that give rise to  non-trivial dynamics in curved spaces\footnote{Considering an inflationary scenario that proceeds towards a final Minkowski minimum, one should distinguish two types of potentials. The first type has a global minimum at a point in the interior of field space. In this case, the global asymptotic stability of the critical point can be proved using La Salle's theorem (with $3H^2$ as the Lyapunov's function). The second type of potentials are positive and vanish only asymptotically at the boundary of the space. In both cases, the monotonicity of the Friedmann constraint implies that the fields will eventually roll towards decreasing values of the potential. Only the second case can result in a scaling solution, with the fields asymptoting towards the Minkowski minimum at the boundary. These global considerations are similar to the single-field case and unaffected by the geometry of the scalar manifold.}.

In full analogy to the single-field case analyzed in Sec.~\ref{sec:singlefield} one can transform the system of Klein-Gordon equations \eqref{kg} as a first order system for the scalar coordinates $\phi^I$ and their (normalised) velocities $v^I\equiv d\phi^I/dN$:
\begin{subequations}
	\label{evfievvi}
	\begin{eqnarray}
	\label{evfi}
	{(\phi^I)'} &=&  v^I \, , \\ \label{evvi}
	(v^I)' &= &-(3 - \epsilon) \left(  v^I  + p^I  \right) - \Gamma^I_{JK} v^J v^K  \,.
	\end{eqnarray}
\end{subequations}
This includes the effect of a curved manifold and hence  velocity-dependent ``centrifugal'' forces via the Christoffel symbols, as well as of the potential energy gradient via  
\beqn
p_I = {\partial (\ln V) \over \partial \phi^I} \,.
\eeqn
Finally, the slow-roll parameter in this case is defined in terms of the velocities as
\beqn
\epsilon =  \tfrac12 v_I v^I  = {3 K \over K+V} \,, \qquad  K = \tfrac12 {\cal G}_{IJ} \dot\phi^I \dot\phi^J \,,
\eeqn
where it is easy to see that $0 \leq \epsilon \leq 3$ again. Contracting equation \eqref{evvi} with $v_I$ gives an evolution equation for $\epsilon$:
\begin{equation} \label{deps}
\epsilon' = -(3 - \epsilon)  \left( 2 \epsilon  +  p_I v^I \right)\, .
\end{equation}
Discarding the kinetic solution with $\epsilon = 3$ (for now), a scaling solution satisfies
\begin{align}  \label{cp_eps1}
\epsilon = - \tfrac12 p_I v^I  \, .
\end{align}
Note that this implies $v_I \mathcal{D}_t v^I = 0$ and hence requires the covariant acceleration  to either vanish or to be orthogonal to the velocity. Moreover, as  $V'/ V = p_I v^I =-2\epsilon$, it also implies that for  a constant $\epsilon$ the evolution of the potential energy is trivially solved to be
\beq
V(N)=V_0e^{-2\epsilon N} \,.
\eeq 
Therefore the potential, if bounded from below (to ensure that the transformation $t \rightarrow N$ is well-defined), should be non-zero and tend to zero at the boundary of the field-space (and not contain any critical points elsewhere).  

The relevance of scaling solutions in inflation becomes apparent if we examine the variation of $\epsilon$. For multiple fields $\phi^I$ and a general field-metric ${\cal G}_{IJ}$, Eq.~\eqref{doteps} generalizes to
\begin{equation}
\dot \epsilon = {\dot \phi_I {\cal D}_t \dot \phi^I  \over H^2} + 2 \epsilon^2H \, .
\end{equation}
A possible generalization of the single-field slow-roll behaviour  is the slow-roll slow-turn approximation \cite{GrootNibbelink:2000vx,GrootNibbelink:2001qt,Peterson:2010np,Peterson:2011ytETAL}, valid when $D_t \dot \phi^I\ll 1$ and $\epsilon \ll 1$.  If the fields are canonically normalized then we can relate velocities to gradients of the potential by requiring a slowly varying $\epsilon_V \equiv \tfrac12 p_I p^I$: 
\begin{equation} \label{mfsr}
{\ud  \epsilon_V \over \ud N} =  {\ud \phi^K \over \ud N}\left( {V^{,I} V_{,IK} \over V^2} - {V^{,I} V_{,I} V_{,K} \over V^3} \right) =   {\ud \phi^K \over \ud N}\left( \eta_K - 2 \epsilon_V p_{K} \right) 
\, ,
\end{equation}
with the definition 
\begin{equation}
\eta_K = {V^{,I} V_{,IK} \over V^2} \, .
\end{equation}
Since ${\ud \phi^K / \ud N}$ can be of order one then $\epsilon_V, \eta_K $ must be small (as also found in Refs.~\cite{Lyth:2009zz,Yang:2012bs}). When this holds, the multifield slow-roll slow-turn approximation is close to a scaling solution of a product-exponential potential in which every field contributes with a non-zero velocity.  However, as we discussed, non-trivial geometries can lead to a departure from the slow-roll slow-turn behaviour, as we will see explicitly.

\subsection{Stability criteria of the background }
\label{sec:adiabatic_entropic}

For later reference it is useful to review the notion of stability on a multi-dimensional curved scalar manifold. 
A minimal requirement for the existence of an ``attractor'' solution $\Phi^I_{sol}$ is the decay of small perturbations around that particular solution, which is determined by the local Lyapunov exponents of the Jacobian matrix evaluated at the solution
\begin{equation} \label{eq:lin}
	(\delta x^I)' = J^I_J \delta x^J \, .
\end{equation}
If every exponent is negative then the system converges to $\Phi^I_{sol}$. The case of zero eigenvalues is more intricate because stability will be determined by higher order terms. Additionally, a positive exponent indicates an unstable solution (with possibly chaotic behaviour). An important note is that the linearized stability gives information only at a given point where the Jacobian matrix is evaluated. Thus, a small perturbation will not evolve according to the linearized equation \eqref{eq:lin}; the Jacobian encodes information only for an infinitesimally small deformation, not defined a priori. The previous imply that an unstable solution can manifest an almost oscillating behaviour because higher order terms can have an opposite effect compared to the leading ones. 

Likewise, stability of fluctuations is treated in the adiabatic/entropic decomposition. For two fields one can distinguish the adiabatic direction, which proceeds along the background inflaton trajectory, and the entropic (or isocurature) direction which is perpendicular to it. The two directions can be defined through the unitary vectors $\hat \sigma^I$ and $\hat s ^I$ 
\begin{align}
\hat\sigma^I &\equiv {\dot \phi^I \over\dot \sigma} \,, \qquad
\hat s^I \equiv {\omega^I\over \omega}
={1\over \sqrt{\cal{G}} }\epsilon^{IJ} \hat \sigma_J 
\,  ,
\end{align}
where $\dot \sigma^2 \equiv {\cal G}_{IJ} \dot \phi^I \dot\phi^J$ is the velocity of the background motion, $\omega^I ={\cal D}_t\hat \sigma^I$ is the turn-rate vector and $\omega = \left | \omega^I \right |$. It is easy to see that both vectors are normalised to unity and perpendicular to each other. Using Eq.~\eqref{deps} we can rewrite the turn rate as
\begin{equation}
\omega^2 = {{\cal G}^{IJ}V_{,I} V_{,J} -V_{\sigma}^2 \over \dot{\sigma}^2} = H^2 (3 - \epsilon)^2 \left[ { \epsilon_V \over \epsilon} - \left({\eta \over 2 (3-\epsilon)}  +1 \right)^2 \right] \,.
\label{eq:omegaepsilonepsilonv}
\end{equation}
When the slow-roll conditions hold, $\epsilon,\eta \ll 1$, the previous reduces to the expression given in \cite{Hetz:2016ics}. We observe two limiting cases: in the slow-turn limit (i.e. $\eta,(\omega/H) \ll 1$) the potential and Hubble slow-roll parameters coincide $\epsilon  \approx \epsilon_V$, whereas in the large turn-rate limit ($\eta \ll 1$ and $\omega/H \gg 1$) $\epsilon \ll \epsilon_V$. Thus, in multi-field inflation with $\epsilon' \ll 1$ we have $\epsilon \leq \epsilon_V$. Projecting the mass matrix 
\beq
\label{eq:MIJfull}
M^I_{~J}={\cal{G}}^{IK} {\cal D}_K {\cal D}_J V-R^I{}_{KLJ}\dot \phi^K \dot\phi^L \,,
\eeq
along these two directions we define the adiabatic and isocurvature (or entropic) components respectively (see for example Refs.~\cite{Kaiser:2012ak, Peterson:2010np, Peterson:2011ytETAL, Renaux-Petel:2015mga, Achucarro:2012sm, Achucarro:2014msa, Gong:2011uw}). The latter is given by $M_{ss} = \hat s_I \hat s^J M^I_{~J}$.
For more than two fields the analogous adiabatic-entropic decomposition proceeds similarly and can be found for example in Ref.~\cite{Kaiser:2012ak}.

An important remark that has been missed so far in the literature of multi-field models concerns the stability criteria for the background. More precisely,  the local Lyapunov exponents will not necessarily involve the effective mass on super-Hubble scales \cite{Christodoulidis:2019mkj}, defined from the isocurvature perturbations equation of motion for $k \ll aH $ 
\begin{equation} \label{meff}
\ddot{Q_s} + 3H \dot{Q_s} + \mu^2_{s}Q_s = 0 \,, \qquad 
\mu^2_{s} = M_{ss} + 3 \omega^2   \, ,
\end{equation}
evaluated along the background solution. As we will outline with specific examples in Secs. \ref{sec:3dcrit} and \ref{sec:4dcrit}  the condition $\mu^2_s>0$ is neither necessary nor sufficient for the existence of an ``attractor solution'' of the zeroth order Klein-Gordon equation (while it is of course necessary for the stability of the full quantum mechanical system). This quantity becomes relevant at first order because it dictates the growth/decay of normalized orthogonal cosmological perturbations. However, in order to substitute the background solution into the linearized cosmological perturbations one needs to provide either an analytical solution or a late time solution (by means of stability analysis) to ensure that this is the correct behaviour of the FLRW background and this can not be done showing $\mu^2_s>0$ for a given background. We can illustrate this by means the following example. Defining linear combinations of the perturbations $y^I\equiv f^I_K \delta x^{K} $ we can write the evolution equation of this variable as
\begin{equation} 
(y^I)' =  \tilde{J}^I_{~K}  y^K \, .
\end{equation}
The sign of the eigenvalues of the new Jacobian matrix provide no information about the system \eqref{eq:lin} and this is roughly what happens when someone infers stability through the adiabatic/entropic split.

Background stability does not imply stability of fluctuations, of course. In a negatively curved manifold, the curvature term in  Eq.~\eqref{eq:MIJfull} can destabilize fluctuations which would be stable in the absence of geometry  \cite{Renaux-Petel:2015mga, Renaux-Petel:2017dia, Cicoli:2018}. The analogous quantity 
 \begin{align}
  m_s^2 = 
  \mu_s^2 - 4 \omega^2 \,,
 \end{align}
governs the evolution of sub-horizon fluctuations and is referred to as the sub-horizon mass of entropic perturbations. 
Given the parametric relation between the two, one can imagine that different relations between $m_s$, $\omega$ and $H$ can lead to different  fluctuation dynamics. 
Furthermore, the sub-horizon evolution of fluctuations for multi-field models with large turn-rate (``strongly coupled'' fluctuations) can have interesting phenomenology (see e.g. Ref.~\cite{Cremonini:2010ua, Achucarro:2016fby, Fumagalli:2019noh,Achucarro:2019pux}).

\subsection{Scalar geometry}
\label{sec:mf2}

For the purpose of clarity, we restrict ourselves to a two-dimensional field-space manifold, which is sufficient to illustrate the effects of a non-trivial scalar geometry and easy to visualize. Moreover, we will mainly consider geometries with a transitively acting isometry (and comment on its generalizations where possible). 
The most general metric with a transitively acting isometry can  be brought to the form
\begin{equation} \label{eq:isom}
\ud s^2 = \ud \rho^2 + f^2(\rho) \ud \phi^2 \, ,
\end{equation}
where the isometry corresponds to shifts in $\phi$ (see App.~\ref{app:isom}). 
Simple examples include the polar coordinates of flat space with $f=\rho$ as well as a parametrization of hyperbolic space of the form  $f = L \sinh(\rho / L)$. While both of these have a vanishing $f$ at $\rho = 0$, this is not necessarily the case: the function $f$ could be nowhere vanishing. Again the hyperbolic space provides an example of such a parametrization, with $f =  \exp(\rho / L)$ spanning the Poincare half-plane\footnote{The Poincare disc parametrization can be brought to the form of \eqref{eq:isom} by a simple coordinate transformation.}. 
In analogy with the definition of the gradient $p_I$ of the scalar potential, we can define $L_\rho$ as 
\beq
{1\over L_\rho} = {f_{,\rho}\over f} \,.
\label{eq:Ldef}
\eeq
The subscript $\rho$ is intended to imply that $L_\rho$ will generically be $\rho$-dependent. A special case in the classification of critical points will be that of $L_\rho$ constant, playing the analogous role of an exponential potential.

In view of the different possibilities, we will not commit to any particular  geometrical interpretation of the two field coordinates. While $\rho$ and $\phi$ can be seen as a radius and angle in certain cases, this would suggest that $\rho$  must have an origin and $\phi$ must have a periodic identification. These requirements are by no means necessary for the ensuing discussion. Alternatively, one can also see the above metric as a fibre bundle, with $\rho$ being the base coordinate and $\phi$ being the fibre. Finally, in the case of a hyperbolic manifold, this system forms a complex axion-dilaton scalar with $\tau = \phi + i e^{-\rho / L}$. We will use this terminology interchangeably in what follows. 

The non-zero Christoffel symbols associated to the above metric are given by  
\begin{equation}
\Gamma^{\rho}_{ \phi \phi}= - {f^2  \over L_\rho} \,, \qquad  \Gamma^{\phi}_{ \rho \phi} =  {1 \over L_\rho} \,,   
\end{equation} 
and the field-space Ricci scalar is
\beq
{\cal R} = - 2{f_{,\rho\rho}\over f} \, .
\eeq
A simple example would be the flat geometry with $f(\rho) = \rho$, which also provides  geometric centrifugal forces as the angular evolution does not proceed along a geodesic. With $f_{,\rho\rho} = 0$, it can be seen as the dividing case between positively and negatively curved manifolds, with concave and convex functions $f(\rho)$ respectively. For instance, a metric function $f=\ln \left( 1 + \rho \right)$ has a Ricci scalar given by
\begin{equation}
{\cal R}= {2 \over (1+\rho)^2\ln(1+\rho)} \, ,
\end{equation}
that is always positive in the domain of definition  $\rho \in (0,+ \infty)$. Similarly, convex functions give rise to negative field-space curvature. An important case is when $f$ is given by any linear combination of exponentials $L e^{\pm \rho / L}$, leading to constant negative curvature and corresponding to different parametrizations of hyperbolic manifolds. Of these, the case with a single exponential, i.e.~the Poincare half-plane, will play a special case in the classification of critical points and can be seen as the field-space equivalent of an exponential potential term.

For the above geometries, the Hubble flow parameter is  
\begin{equation}
\label{eq:epsilonfrho}
\epsilon = {1 \over 2} \left[(v^{\rho})^2+f^2(v^{\phi})^2 \right] \,.
\end{equation} 
We will focus on those critical points for which both terms are separately constant. This requires both the velocity $v^{\rho}$ as well as the combination  $f v^{\phi}$ to be constant. The latter has the geometric interpretation of being the projection of the general velocity vector $v^I$ along the normalised Killing direction, $y = k_I v^I$, which is given by $k_I = (0, f)$. Similarly, projecting the velocity along the unit vector $b^I$ orthogonal to the Killing vector, $k^I b_I=0$, we obtain $x\equiv b_I v^I$. Introducing the coordinates $(x,y) \equiv (v^{\rho}, fv^{\phi})$ for these components, the Hubble flow parameter becomes
\beq
\epsilon = {1 \over 2} \left (x^2+y^2\right)  \, ,
\eeq
and we will focus on the cases where $x'=y'=0$. These will have $\epsilon'=0$ and hence will correspond to scaling solutions.

\section{Scaling solutions for systems with an integral of motion
} 
\label{sec:3d}

\subsection{Critical points analysis} \label{sec:3dcrit}

We are now ready to start classifying the possible critical points for different parameter values. Given that we have fixed the metric, the classification will be based on which $p_I$'s are non zero.

As a trivial first case we consider a cosmological constant ($p_I = 0$). In this case, both velocities will be zero and the scaling solution describes De Sitter space. It is the late-time attractor of the system since contraction with $\dot \phi_I$ in the Klein-Gordon equations \eqref{kg} with a constant potential gives the evolution equation for the kinetic energy in terms of the e-folding number:
\beq
\dot K + 3 H K =0  \qquad \Rightarrow \qquad
K(N)=K(0)e^{-3N} \, .
\eeq
The kinetic energy of every field vanishes asymptotically and it makes no difference if we include non-trivial dependence on the metric, because $\epsilon$ is still converging to zero. The future asymptotic state of the system is a De Sitter space whereas the kinetic solution is excluded because it corresponds to infinite initial kinetic energy. 

A more interesting case is a non-trivial scalar potential that still preserves the isometry of the metric, i.e.~that has a shift symmetry in the potential along the $\phi$ direction. This requires $p_{\phi} = 0$ and there is only a non-zero gradient along the $\rho$ direction. In this case, there is one integral of motion following from the shift symmetry, which if one thinks of $\phi$ as an angular variable is the conservation of angular momentum
\begin{equation}
\pi_{\phi}=a^3 f^2 \dot \phi \, ,
\end{equation}
and only three out of the four variables $\phi^I,v^I$ are independent; the state space is 3-dimensional and the dynamical system takes the simple form
\begin{subequations}
	\label{dx3D}
	\begin{eqnarray}
	x' &=& - (3-\epsilon)  \left(x + p_{\rho}  \right) + { y^2  \over L_\rho }  \,, \\ 
	y' &=& - \left (3-\epsilon     +  { x   \over L_\rho }  \right )y\,.
	\end{eqnarray}
\end{subequations}
Critical points of this system will have $\epsilon'=0$ and hence will correspond to scaling solutions.  
Eq.~\eqref{cp_eps1}  for constant $p_\rho$ implies that both $x$ and $y$ are constants on the scaling solution. There are three solutions in this case, given as critical points of Eq. \eqref{dx3D}.
\begin{itemize}
	\item The first scaling solution has vanishing angular velocity and hence 
	takes places along the radius-like
	$\rho$ coordinate, with the critical point given by 
	\begin{equation} \label{gradient}
	(x,y)_{\rm grad} = (- p_\rho,0) \,, 
	\end{equation}
	leading to $\epsilon = p^2_\rho/2$ and hence exists only for $p_\rho<\sqrt{6}$.
In this case the velocity points along the gradient direction; the field slides down the scalar potential along the radial coordinate. Indeed this proceeds along a geodesic: both the potential and the centrifugal forces in the orthogonal direction are zero. It resembles the flat metric in the sense that $\epsilon$ receives contribution only from the $\rho$ field, while any initial velocity of the second field is redshifted away, and it is given by the projection of the gradients along the direction orthogonal to the Killing vector: $\epsilon= (b^I p_I)^2/2$, which coincides with $\epsilon_V$, resulting in zero turn rate.
	
	\item For a hyperbolic space with $f = e^{\rho/L}$, one can have a non-vanishing angle-like velocity
	with solution\footnote{For an exponential metric function $f$ the definition of Eq.~\eqref{eq:Ldef} leads trivially to $L_\rho=L$. We will however use $L_\rho$, as it makes the connection to more general cases and approximate scaling solutions more transparent. }
	\begin{align} \label{hyper}
	(x,y)_{\rm hyper}  &= \left( - {6 \over {2 \over L_\rho} + p_\rho}, \pm{\sqrt{6} \sqrt{p_\rho^2 + 2{p_\rho \over L_\rho} -6}  \over {2 \over L_\rho} + p_\rho} \right) \, .
	\end{align}
In this picture the background motion corresponds to a spiralling behaviour, where the field moves in the angular direction $\phi$ and has a decreasing radial value $\rho$. The inflationary trajectory therefore does not follow the potential gradient, nor is it along a geodesic. 
This solution exists provided 
	\beq
	\label{critical}
	p_\rho > p_{\rm crit} = -{1 \over L_\rho} + \sqrt{6+ {1 \over L_\rho^2}} \, ,
	\eeq
which defines a critical value for the  gradient of symmetric potentials. The Hubble flow parameter for this solution can be given by Eq.~\eqref{eq:epsilonfrho} as
\beq
\label{eq:epsilonspiralling}
\epsilon = {3 L_\rho p_{\rho} \over 2+L_\rho p_\rho} \, .
\eeq

\item Finally, there are two kinetic solutions $(x,y)_{\rm kin} = (\pm \sqrt{6},0)$.
	\end{itemize}
We next turn to stability of these critical points. For the hyperbolic space with an exponential function $f$, there exists an invariant two-dimensional subspace spanned by $x$ and $y$. The asymptotic behaviour will be determined by the system's critical points and their local stability. The Jacobian matrix evaluated at the gradient solution has eigenvalues 
 \begin{align}
   ( \lambda_{1}, \lambda_2)_{\rm grad} = \left(-{1 \over 2} \left( 6-p_\rho^2 \right), -3 + {p_\rho \over L_\rho} + {p_\rho^2 \over 2}\right) \,, 
   \label{eq:3Deigenvaluesgradient}
\end{align} 
which are both negative when $p_\rho < p_{\rm crit}$. The gradient solution becomes unstable when the   gradient along the $\rho$ direction exceeds the critical value, where the hyperbolic solution appears. The eigenvalues for the hyperbolic solution are more complicated but they always have a negative real part
whenever $p_{\rho}>p_{cr}$ and $L_{\rho}>0$ (see App.~\ref{app:stab}). Hence, when the hyperbolic   solution exists, it is always stable.

\begin{figure}[t!]
	\centering
	\includegraphics[width=1\textwidth]{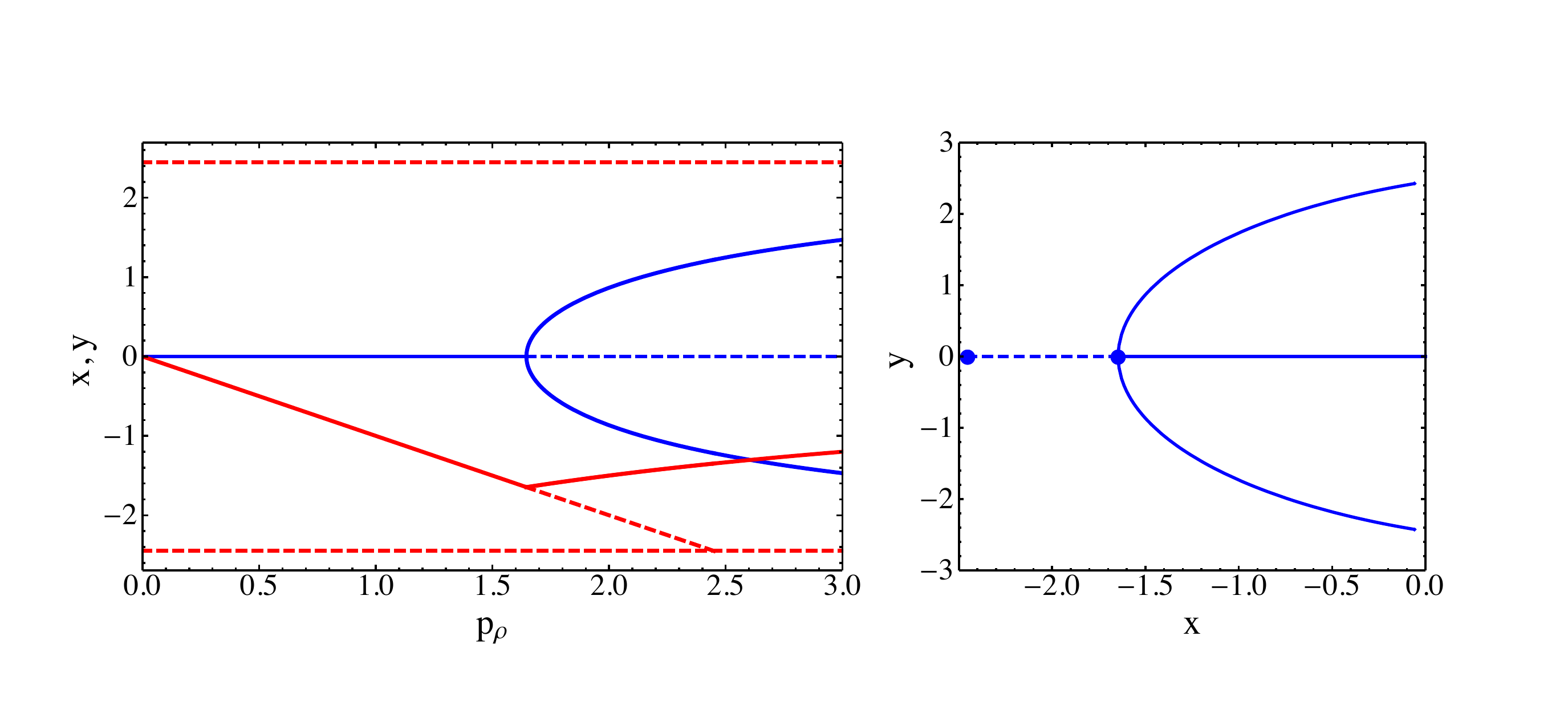}
	\caption{
{\it Left:} The bifurcation diagram for the system of Eqs.~\eqref{dx3D} on a hyperbolic space $f=L e^{\rho/L}$ for a shift symmetric potential with $p_\phi=0$. The red and blue curves correspond to the values of $x$ and $y$ respectively, as a function of $p_\rho$, at the various critical points. Solid (dashed) lines correspond to stable (unstable) solutions. We see that  for large values of the potential steepness $p_\rho$ the only stable solution is Eq.~\eqref{hyper}. 
{\it Right:} The same bifurcation diagram given as a curve on the $(x,y)$ plane.	
}	\label{fig:bifurcation}
\end{figure}

\begin{figure}[t!]
	\centering
	\includegraphics[width=0.45\textwidth]{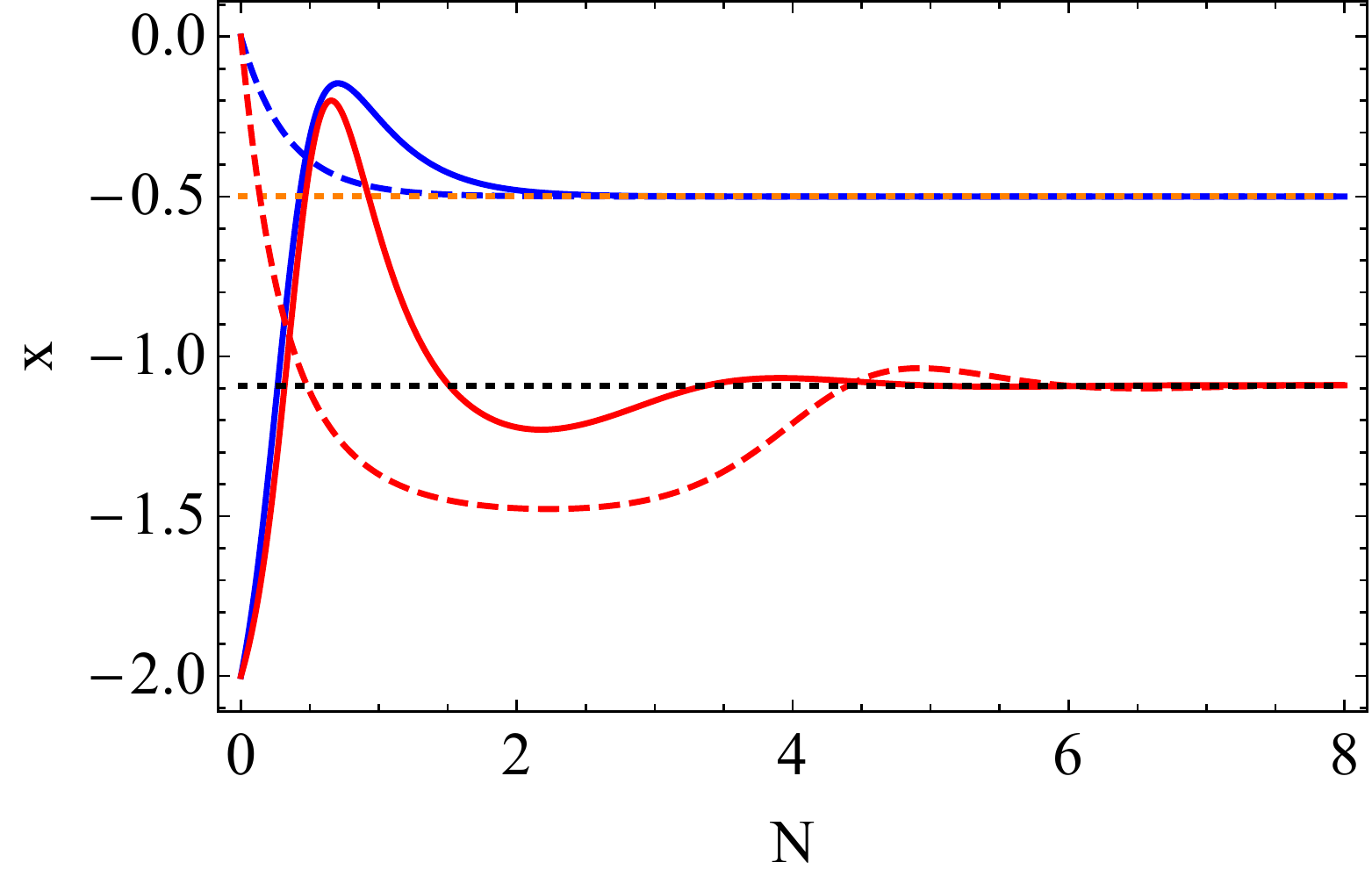}
	\includegraphics[width=0.45\textwidth]{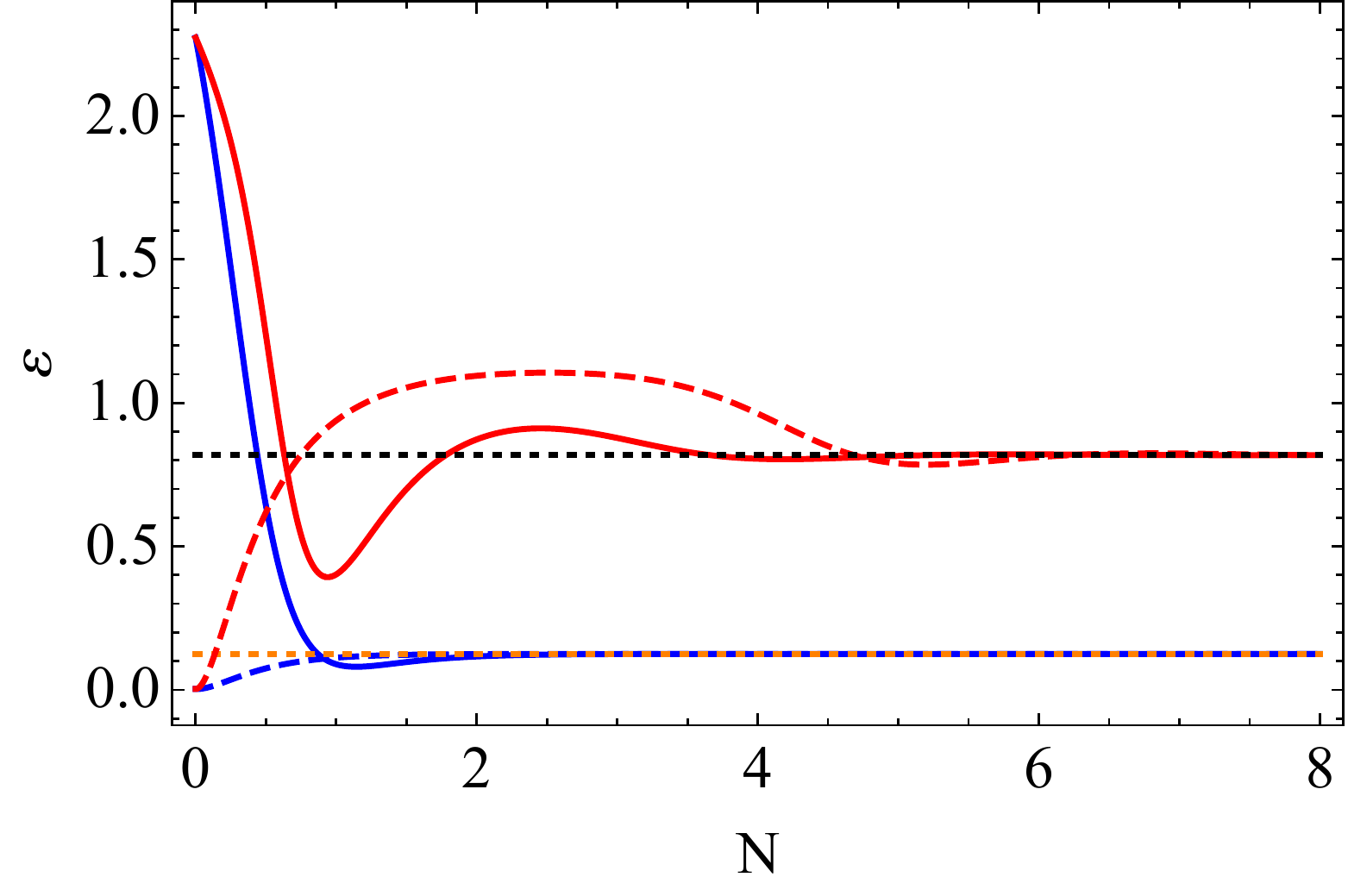}
	\caption{{\it Left:} Evolution of $x$ for the hyperbolic space with $L=0.5$ and an exponential potential for $p_\rho=0.5$ (blue) and $p_\rho=1.5$ (red). Solid lines correspond to initial conditions close to kinetic domination, while dashed lines close to potential domination. These sub- and super-critical  values for $p_\rho$ give rise to the gradient  (orange dotted) and hyperbolic (black dotted) solution, respectively. {\it Right:} Evolution of $\epsilon$ for the same parameters and color coding.  
	}
	\label{fig:3d}
\end{figure}

The behaviour of these critical points is illustrated in Fig.~\ref{fig:bifurcation}, showing the bifurcation diagram of the system in the case of a hyperbolic metric with $f = e^{\rho/L}$. We see that the curve $y(p_\rho)$ has the typical form of a supercritical pitchfork bifurcation. The branch $y=0$ is stable for $p_\rho<p_{cr}$ and becomes unstable for $p_\rho>p_{\rm crit}$. The total number of stable minus unstable critical points is conserved through the creation of two stable branches (blue solid curves) for $p_\rho> p_{\rm crit}$, given by Eq.~\eqref{hyper}. 
The approach to the various critical points in the case of a hyperbolic metric is numerically illustrated in    Fig.~\ref{fig:3d}. 
Therefore, when $p_{\rho}$ and $L_{\rho}$ have the same sign kinetic solutions are unstable for the hyperbolic space and never form an asymptotic attractor.

It is interesting to compare the value of $\epsilon$ for the geodesic and hyperbolic solution. By comparing Eq.~\eqref{eq:epsilonspiralling}, which holds for the hyperbolic solution, to $\epsilon=p_\rho^2/2$, which holds for the geodesic solution, we can immediately see that $\epsilon_{\rm hyper}< \epsilon_{\rm grad}$. Solving this condition with respect to $p_\rho$ we recover $p_{\rm crit}$ appearing in Eq.~\eqref{critical}. Hence, whenever both solutions are present, the system will dynamically ``choose'' the one with smaller $\epsilon$. Referring to Eq.~\eqref{eq:omegaepsilonepsilonv}, we see that between the two solutions, the one with smaller $\epsilon$ corresponds to a solution with a non-zero turn rate $\omega^2$.

For more general metrics - which  are not described by an exponential function $f(\rho)$- the subspace includes $(\rho,x,y)$, and since $x\neq 0$ there are no critical points in the usual sense. Nevertheless, the $(x,y)$ subspace admits the gradient solution Eq.~\eqref{gradient}, in addition to the kinetic solutions $x = \pm  \sqrt{6}$. The stability of the gradient solution of Eq.~\eqref{gradient} in the case of a general metric function with non-constant $L_\rho$ is determined by the eigenvalues of the stability matrix, which correspond to local Lyapunov exponents.  They are given again by the eigenvalues of Eq.~\eqref{eq:3Deigenvaluesgradient}, with the addition of a third eigenvalue $\lambda_3=0$ (a reflection of our ability to define the gradient solution for every $\chi$). The gradient solution is thus stable for $p_\rho< p_{\rm crit}$, with $p_{\rm crit}$ given by the expression of Eq.~\eqref{critical}, albeit with $L_\rho$ not being constant in the case of a general non-exponential metric function $f(\rho)$\footnote{If $p_\rho > p_{\rm crit}$ then solutions with $y=0$ (gradient or kinetic) are unstable. This is possible if $p_\rho L_\rho >0 $ and so the ``centrifugal force'' can balance the gradient term, leading to a situation where the system slowly approaches a De-Sitter phase: $x\rightarrow 0$ and $y\rightarrow 0$.}.  

It is worth noticing that for a hyperbolic space $ (\mu_s/H)^2=-p_{\rho}/L_{\rho}\lambda_{2,\rm grad}$ and so positivity of the effective isocurvature mass is equivalent to a negative eigenvalue $ \lambda_{2,\rm grad}$. This is no longer true for any other $f$ because the second derivative is not a constant multiple of the first derivative. Thus, we can construct models which have a stable background while the $k=0$ mode of the perturbation equations is unstable and vice versa; this behaviour was first observed in \cite{Cicoli:2018}. As an example, consider $f=e^{\rho^2}$ and a positive $p_{\rho}$. In this case, when $\rho$ is negative $p_{crit}$ will always be greater than $\sqrt{6}$ and so the solution will be stable according to the analysis of this section. On the other hand, the effective mass, evaluated along the gradient solution, is equal to
\begin{equation}
\left({\mu_s \over H}\right)^2  = 2 p \left[\rho (3-p^2/2)- p(2\rho^2 + 1)\right] \, .
\end{equation}
The contribution from the negative curvature will always dominate at late times and so after a certain number of e-folds $\mu^2_s$ will become negative, although the solution to the zeroth order Klein-Gordon equation is stable. Physically, this indicates that the proper distance between different late-time trajectories diverges.

\subsection{Comparison to inflationary models}
\label{sec:3Dcomp}

An example that is close to a scaling solution of this class has been coined   ``hyperinflation'' \cite{Brown:2017osf} and was studied in e.g.~\cite{Mizuno:2017idt, Bjorkmo:2019aev}. The field-space has constant negative curvature, and is written in global coordinates as
\begin{equation}
	\ud s^2 = \ud \rho^2 +  L^2 \sinh^2\left({ \rho \over L} \right)\ud \phi^2 \, ,
\end{equation}
where $\rho, \phi$ are radial and angular-like variables respectively. Note that the metric function $f$ asymptotically behaves as an exponential for $\rho \gg L$. The late-time solution of equation \eqref{hyper} for this choice of metric can be written as
\begin{equation}
	\dot{\rho}   \approx - {3 H L \over   \coth \left({\rho \over L}\right) +  {1 \over 2} p_\rho  L } \, ,
\end{equation}
where we took the local (field-dependent) value of $L_\rho$ to be defined according to Eq.~\eqref{eq:Ldef}
\beq
L_\rho  = L \tanh \left ( {\rho \over L} \right ) \, ,
\eeq
and the radial velocity to be $\dot \rho = H \rho' \equiv H  x$. For large field-space curvature $L \ll 1$ at $\rho \gg L$ the second term in the denominator is subdominant and thus the radial velocity becomes $\dot{\rho}  \approx - 3 H L$ which is the ``attractor'' expression given for this model in Refs.~\cite{Brown:2017osf,Mizuno:2017idt}. Moreover, the steepness condition for the potential $p_{\rho}>3L$ corresponds to the small $L$ limit of Eq.~\eqref{critical}.

The simplest possible single-parameter potential given by $V(\rho)=m^2 \rho^2$ provides an interesting illustration of the bifurcation phenomenon, given a hyperbolic metric. The potential steepness is $p_\rho=2/\rho$, which is a monotonically decreasing function for $\rho>0$. 
 At large field values, the gradient $p_\rho$ becomes arbitrarily small and hence the gradient solution is stable. Upon rolling down to smaller values of $\rho$, the gradient increases and will hit the critical value $p_{\rm crit}$ around $\rho \sim 1/L$. At this point there will be a transition to the hyperbolic solution. 
For an exponential metric function $f(\rho) = L e^{\rho/L}$, inflation will ultimately end when the fields approach the minimum of the potential. In the case of a different parametrization of hyperbolic space, like the one used above $f(\rho)=L\sinh(\rho/L)$, the hyperbolic solution will continue up to around $\rho \sim L$ when the exponential approximation of the metric ceases to be valid. For small $L$, corresponding a highly curved field-space manifold, a prolonged period of hyperbolic inflation will exist (see Ref.~\cite{Bjorkmo:2019aev} for a recent discussion of hyperinflation and Ref.~\cite{Christodoulidis:2019mkj} for an illustration of the bifurcation in the case of a quadratic potential.). 

To illustrate this interplay between the different critical points, we  analyse the implications of the bifurcation diagram of Fig.~\ref{fig:3d} in a simpler set-up that will be closer to the present scaling discussion. We saw in Sec.~\ref{sec:singlefield} that models like quadratic and Starobinsky inflation can be approximated by scaling solutions with field-dependent steepness parameter $p$. We can analyze two-field models in the same way: let us consider a hyperbolic space $f=Le^{\rho/L}$ with a shift symmetric potential in $\phi$ with a field-dependent gradient $p_\rho = p_\rho (\rho)$. A simple example is given by 
 \beq
  p_\rho = p_0 +   \Delta p  \tanh\left (  \alpha \rho \right) \, .
 \label{eq:steepnessvar}
 \eeq
arising from the potential 
\beq
  V(\rho) = V_0 \,e^{p_0 \rho} \left [\cosh(\alpha \rho)\right]^{\Delta p \over \alpha} \, .
\eeq
Note that this simply amounts to a sum of exponentials for the case of $\alpha=1$ and $\Delta p =1$. In addition, for a range of hyperbolic curvatures, this potential crosses the critical gradient around the origin: for $\alpha>0$ the potential steepness $p$ exceeds the critical value $p_{cr}$ for large field values $\rho$, while for $\alpha <0$ the opposite occurs, $p_\rho<p_{cr}$ for $\rho \to \infty$. 

\begin{figure}[t!]
	\centering
	\includegraphics[width=0.48\textwidth]{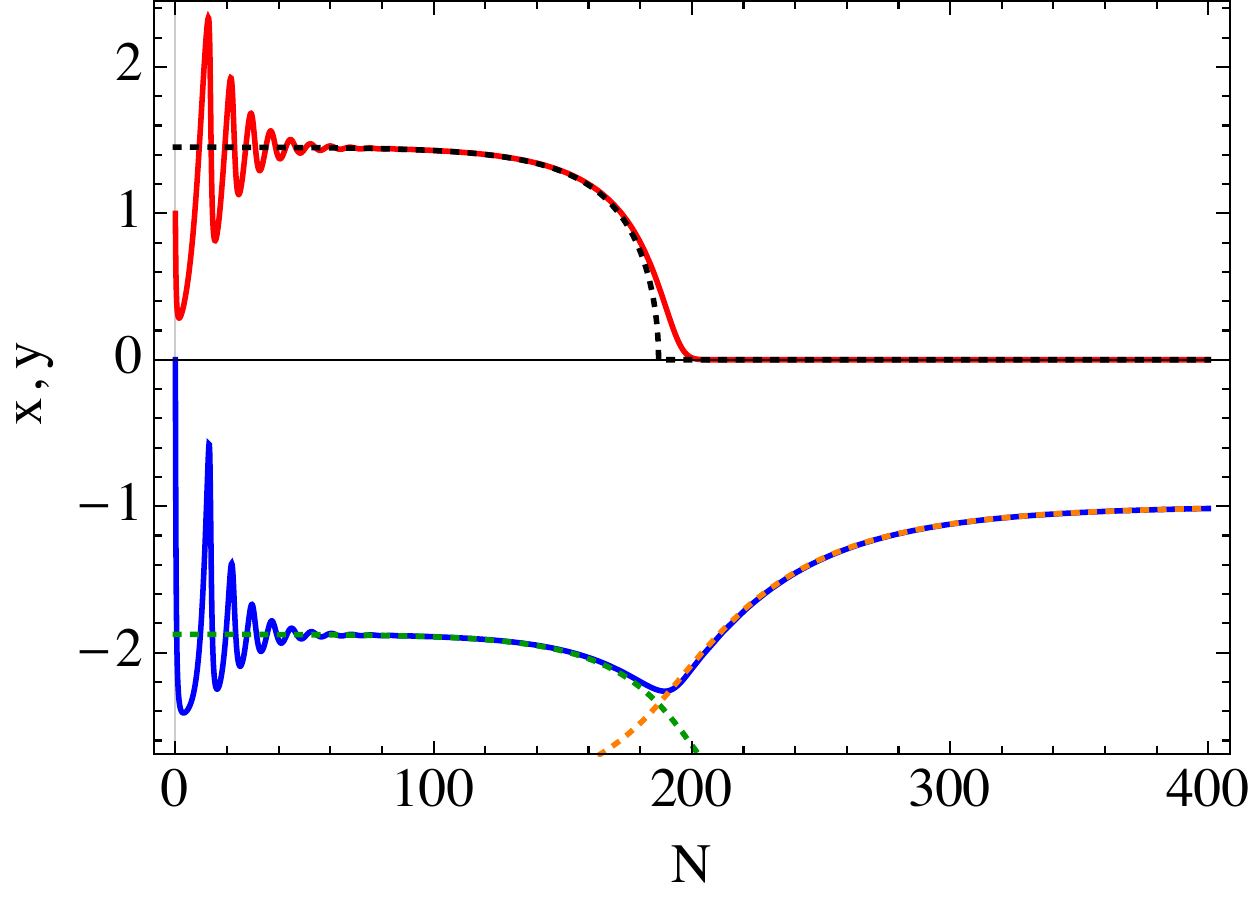}
	\includegraphics[width=0.48\textwidth]{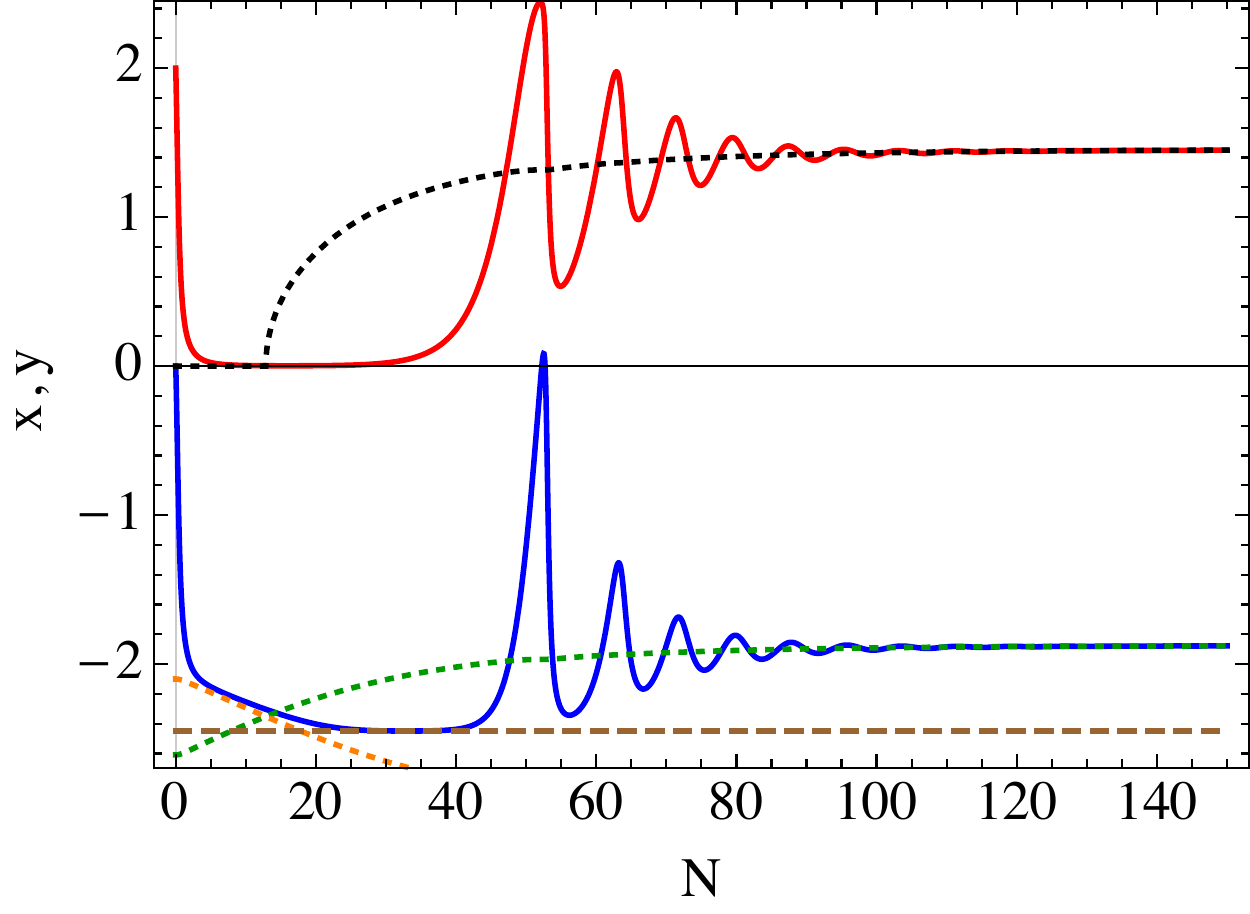}
	\caption{{\it Left:} Evolution of $x$ and $y$ (blue and red respectively) for the potential steepness of Eq.~\eqref{eq:steepnessvar} with $\alpha>0$ (left) and $\alpha <0$ (right). The dotted curves correspond to the scaling solutions evaluated at $p_\rho=p_\rho(\rho)$. The brown dashed line on the right panel corresponds to the kinetic solution $x=-\sqrt 6$. The parameters used are $L=10$ for the field-space and $|\alpha|=0.01$, $p_0=2$ and $\Delta p =1$ for the potential.
	}
	\label{fig:SSB}
\end{figure}

Fig.~\ref{fig:SSB} shows a characteristic field evolution in both cases $\alpha >0$ and $\alpha <0$ for $L=10$ and hence $p_{\rm crit} \sim 2.35$, $|\alpha|=0.01$,  $p_0=2$ and $\Delta p =1$.  In the former case (left panel of Fig.~\ref{fig:SSB}), the steepness of the potential at the initial field value is supercritical, so the system quickly (after some transient oscillations) follows the hyperbolic solution of Eq.~\eqref{hyper}. As the system evolves towards smaller values of $\rho$, the potential steepness reduces and the hyperbolic solution smoothly ceases to exist, as shown in Fig.~\ref{fig:bifurcation}. From this point onwards the system evolves along the subcritical gradient solution \eqref{gradient}.

In contrast, for the case $\alpha <0$ (right panel of Fig.~\ref{fig:SSB}), the steepness of the potential at the initial field value is subcritical, so the system relaxes to the gradient solution. As $\rho$ rolls towards smaller values, the steepness of the potential increases. Since
$p_\rho > \sqrt{6}$ the system first enters the kinetic regime $x\simeq -\sqrt{6}$, as shown by the brown-dashed line. After some $e$-folds, the $\phi$ field gets de-stabilized and the system follows the hyperbolic solution of Eq.~\eqref{hyper}. The condition $p_\rho > p_{\rm cr}$ occurs at $N\simeq 15$, as shown by the black-dotted line. However $y$ has come exponentially close to zero by the initial phase, so that the destabilization takes several $e$-folds. In reality, one should take into account the quantum fluctuations, which can  de-stabilize the system and push it to the hyperbolic solution earlier. 

Overall, Fig.~\ref{fig:SSB} shows both the transition from hyperbolic to gradient solution, as well as the transition from gradient (or even kinetic) to hyperbolic. The latter is a typical example of spontaneous symmetry breaking, since the equations of motion are invariant with respect to $y\to -y$, whereas the solution picks up a definite sign of $y$, determined by initial conditions or quantum fluctuations. In particular, the approach to a gradient or kinetic solution occurs at an exponential rate for $p_\rho<p_{\rm cr}$ and hence a classical system can be put arbitrarily close to it. In a real scalar field system, quantum fluctuations put a lower limit on how close to  $y=0$ the system can be. Stochastic effects in inflation play an important role in cases of spontaneous symmetry breaking, such as hybrid inflation \cite{Linde:1993cn, Clesse:2015wea, Lyth:2012yp, Guth:2012we}, and have been studied lately using analytical and numerical techniques (e.g. Refs.~\cite{Grain:2017dqa,Pinol:2018euk}). The same behavior depicted in Fig.~\ref{fig:SSB} can also arise in systems with a field-dependent field-space curvature instead of a field-dependent potential steepness, or even in systems where both the field-space curvature $L$ as well as the potential steepness $p_\rho$ are field-dependent.

\section{Scaling solutions for systems with an isometry and generic potentials
} \label{sec:4d}

\subsection{Critical points analysis} \label{sec:4dcrit}

Finally, we turn to the case of potentials that depend on both fields $\rho$ and $\phi$. The dynamical system of Eq.~\eqref{dx4D} is generalized to
\begin{subequations}
	\label{dx4D}
	\begin{eqnarray}
	x' &=& - (3-\epsilon)  \left(x + p_{\rho}  \right) + { y^2  \over L_\rho }  \,, \\ 
	y' &=& - (3-\epsilon)  \left( y + {p_{\phi} \over f } \right)  -  { x y  \over L_\rho } \,.
	\end{eqnarray}
\end{subequations}
Critical points of this system will have $\epsilon'=0$ and hence will correspond to scaling solutions.  

We focus on product separable potentials with either an exponential dependence on $\phi$ or in $\rho$; we have not found any scaling solutions when there is not exponential dependence on either of the coordinates. For such potentials, there are generically no conserved charges and the configuration space is four-dimensional with $(\phi^I, v^I)$ all independent variables\footnote{The case of a possible ``hidden'' conserved charge and its connection to the analysis of Sec.~\ref{sec:3d} is studied separately in Sec.~\ref{sec:hiddenintegralsofmotion}.}.  Since we have assumed non-zero gradients $p_I$, each $v^I$ must become constant (in order for Eq.~\eqref{deps} to contain constant terms) and so their values must be given as critical points of the above dynamical system. 

We start with the case of an exponential dependence on the radius-like coordinate,
 \begin{align}
 \label{eq:frozenpotential1}
\qquad V = e^{p_{\rho}\rho} \tilde h (\phi) \, ,
 \end{align}
and  distinguish the following cases:
	\begin{itemize}
	\item
A solution similar to the gradient solution of the previous section is possible if  the function $h(\phi)$ has an extremum at some value:
\beq
(\phi,x,y)_{\rm grad} = \left( \phi_{\rm grad},-p_{\rho},0 \right ) \, . \label{grad2}
\eeq
This is essentially a single field trajectory, where the turn rate vanishes. It resembles the gradient solution of the symmetric potential; however, in this case inflation proceeds along a valley with $\tilde h$ minimised at some ``angular'' value $\phi_{\rm grad}$.

	\item
	The only other scaling solution is the kinetic one, with $\epsilon = 3$ and $y=0$. 
\end{itemize}
The kinetic solution with $x=\pm \sqrt{6}$ has local eigenvalues of the Jacobian matrix 
  \begin{align}
  \{\lambda_{1,2,3,4}\}_{\rm kin} = \left(0,0,\sqrt{6} \left(\sqrt{6} \mp p_{\rho} \right),{ \pm \sqrt{6} \over L_\rho} \right) \, ,
 \end{align}
and hence there is a stable solution for $p_{\rho} L_\rho < 0$ and $p_{\rho}^2>6$. The gradient solution exists provided $p_{\rho}^2< 6 $ and has local eigenvalues
\begin{equation}
	\{ \lambda_{1,2,\pm} \}_{\rm grad} = \left( 0,-(3-\epsilon), {1 \over 2} \left( A_{\rm grad} \pm \sqrt{A_{\rm grad}^2 - B_{\rm grad}} \right) \right) \, ,
\end{equation}
where
\begin{equation}
\epsilon = {1 \over 2} p_{\rho}^2 \, , \qquad	A_{\rm grad}  =  -(3-\epsilon) + {p_{\rho} \over L_{\rho}} \, , \quad B_{\rm grad} = 4 \left(3-\epsilon \right) { V_{,\phi \phi}(\phi_0)\over f^2 V} \, .
\end{equation} 
$A_{\rm grad}$ is negative for $p_{\rho} < p_{\rm crit}$, where $ p_{\rm crit}$ is given by Eq.~\eqref{critical}, whereas $B_{\rm grad}$ is positive when $\tilde{h}$ has a local minimum at the critical value $\phi_0$. For the former a sufficient condition is $p_{\rho} L_\rho < 0$, in which case $|p_{\rm crit}| > \sqrt{6}$. The stability conditions are therefore identical to those of Sec.~\ref{sec:3dcrit} with the additional requirement $V_{,\phi\phi}>0$. Similar to Sec.~\ref{sec:3dcrit}, if the space is hyperbolic with the specific parametrization $f=e^{\rho/L}$ the effective mass is given by $ \mu_s^2/H^2 = B_{\rm grad}/4 -  p_{\rho}/L_{\rho} A_{\rm grad}$. A positive effective mass requires $B_{\rm grad}> 4 p_{\rho}/L_{\rho} A_{\rm grad}$, which is a weaker condition than the stability requirement resulting from studying the corresponding eigenvalues. Thus, for a potential with a minimum $B_{\rm grad}>0$ and for $p_{\rho}>0$ after some $\rho<0$ the first term of the effective mass will always dominate over the second implying $\mu_s^2>0$. However, if $p_{\rho}$ is supercritical the background trajectory will be unstable, even though the homogeneous mode is stable. The above discussion shows that the background trajectory can be unstable, even if  $\mu_s^2>0$.

The case of an exponential dependence on the angle-like coordinate,
 \begin{align}
 \label{eq:frozenpotential}
\qquad V = h(\rho) e^{p_{\phi}\phi}
 \end{align}
exhibits a richer phenomenology, which falls into different cases:
\begin{itemize}
	\item In order for $y$ to be constant the metric function should also become constant and hence $v^{\rho}$ vanishes, i.e.~this coordinate will be frozen. A critical point with $x=0$ requires the following relation for the $\rho$ coordinate
	\begin{align}
	\label{eq:conditionfrozen}
	f^2 = \frac{p_\phi^2}{6} \left( 1 + \frac{2}{p_\rho L_\rho} \right) \, ,
	\end{align}
	where $L_\rho$, $p_\rho$ and $f$ are generically functions of  $\rho$. This will   fix the coordinate $\rho$ to a specific constant value\footnote{Since $\rho$ is constant throughout the trajectory, the existence of this solution only probes the first derivative of the scalar potential and metric function along $\rho$. This is why the frozen solution does not require $V(\rho)$ and $f(\rho)$ to be exponential in $\rho$, unlike the hyperbolic case of Sec.~\ref{sec:3d}.} $\rho_0$. In the radius ($\rho$) and angle ($\phi$) terminology, the former is constant due to a balance between the radial gradient of the potential and the centrifugal force of the non-geodesic trajectory, while the field spins along the angular direction.
	This is a critical point\footnote{In principle Eq.~\eqref{eq:conditionfrozen} can have an arbitrary number of solutions. We will only consider cases where it has up to two isolated solutions or a continuous set of values, as in shift-symmetric orbital inflation \cite{Achucarro:2019pux}.} with normalised velocities given by
	\begin{align}
	\label{frozen}
	(\rho,x,y)_{\rm froz} = \left (\rho_0 ,0\, ,\, - {p_{\phi} \over f} \right  ) \,.
	\end{align}

In this case, $\epsilon$ is equal to the projection of the gradient along the Killing direction $\epsilon=(k^I p_I)^2/2$, therefore it is always smaller than $\epsilon_V$, resulting into moderate to large turn-rate. Moreover, a noteworthy property of this solution is that $\epsilon$ can be rewritten in a form that is independent of the $\phi$ gradient, resulting in 
\begin{align}
\epsilon = \frac{3 p_\rho L_\rho}{2 + p_\rho L_\rho} \,, 
\label{eps-frozeng}
\end{align}
which is smaller than $\epsilon = 3$ corresponding to the kinetic regime for any finite $p_\rho$. One can immediately note that this is identical to Eq.~\eqref{eq:epsilonspiralling}, derived for the hyperbolic solution with a radial potential. A deeper connection between the two will be uncovered in Sec.~\ref{sec:hiddenintegralsofmotion}.

\item In the case when $h(\rho)$ and $f(\rho)$ each exhibit an extremum at the same value $\rho_{\rm extr}$, a solution similar to the gradient solution emerges, which we will call ``extremum solution''
\beq
(\rho,x,y)_{\rm extr} = \left( \rho_{\rm extr},0,-{p_\phi\over f} \right ) \, ,
\label{eq:extremumsol}
\eeq
for the value $\rho_{\rm extr}$ of the coordinate such that $f$ and $h$ are extremised.
The Hubble flow parameter becomes
\beq
\epsilon = {p_\phi^2 \over 2\, f^2} \, .
\eeq
Note that this is very similar  to the previously mentioned gradient solution \eqref{grad2} with inflation occurring along a geodesic. Furthermore, the extremum of $h$ ensures that there is no radial gradient; hence there is no competition between potential and centrifugal forces as discussed in \cite{Christodoulidis:2019mkj}; instead, both vanish separately.

 \item Finally, if Eq.~\eqref{eq:conditionfrozen} admits an infinite number of solutions where the critical values form a continuous curve then the solution is neutrally stable since the corresponding eigenvalue is zero. 
 
\end{itemize}

 For the frozen solution, the invariant subspace consists of $(\rho,x,y)$ and the eigenvalues of the stability matrix are of the form
\begin{align}
	( \lambda_{1,\pm})_{\rm froz} & = ( -A_{\rm froz}, -{1 \over 2} \left(A_{\rm froz} \pm \sqrt{A_{\rm froz}^2 - B_{\rm froz}} \right) \,.
\end{align}
where $A_{\rm froz}=3- \epsilon >0$. A necessary and sufficient condition for the local stability of the frozen solution is $B_{\rm froz} > 0$.

In the simple case of a product exponential potential and an exponential form of $f$, leading to constant values for $p_{\rho}$ and $L_\rho$, the stability parameter $B_{\rm froz}$
 is given by  
\begin{align}
	B_{\rm froz}=  \frac{48 \epsilon}{(3-\epsilon) L_\rho^2} =   {8 \over L_\rho^2} (2+ p_\rho L_{\rho}) \,.
\end{align}
where we used the Eq.~\eqref{eps-frozeng} for the second equality.
We thus showed that the frozen solution with an exponential potential and an exponential metric function is always stable, and hence this critical point is a dynamical attractor for all parameter values (assuming $p_\rho, L_\rho > 0$).

Linearized stability in the more general case  with non-exponential functions  requires  the  expression
\begin{equation}
\label{eq:generalstabilityconditionfrozeng}
 B_{\rm froz} = - 8 \epsilon \left( {f_{,\rho \rho} \over f} -{9 - \epsilon \over 3 - \epsilon}  \frac{f_{,\rho}^2}{f^2} \right)	 + 4(3-\epsilon ) (p_{\rho})_{,\rho} \, ,
\end{equation}
to be negative. Using the adiabatic/entropic decomposition reviewed in Sec.~\ref{sec:multifield_def}, we can show that the quantity $B_{\rm froz}$ is proportional to the effective mass of isocurvature perturbations on super-Hubble scales. Since motion proceeds along the $\phi$ direction, we have 
\begin{equation}
	\hat\sigma = \left( 0,{\dot{\phi} \over \sqrt{f^2 \dot{\phi}^2}} \right) \, , \qquad \omega^2 = {V_{,\rho}^2 \over \dot{\sigma}^2} =  {2\epsilon H^2 \over L_\rho^2}  \, ,
\end{equation}
and so indeed\footnote{It is a rather straightforward algebraic exercise to relate $B_{\rm froz}$ to $\mu_s^2$, but some components are clearly visible: the term in $B_{\rm froz}$ that includes $f_{\rho\rho}$ is identified to the Riemann term in Eq.~\eqref{eq:MIJfull}. The term in $B_{\rm froz}$ that includes the derivative of $p_{\rho}$ contains the second derivative of the potential, which is the first term in Eq.~\eqref{eq:MIJfull}. The remaining terms can be grouped to give the turn-rate contribution of Eq.~\eqref{meff}. } $B_{\rm froz}= 4 \mu^2_{s}/H^2$. Note that the equivalence of the $\mu_s^2>0$ to the stability criterion derived from the Lyapunov analysis of the background dynamical system arises for a trajectory that proceeds along an isometry of the field-space manifold.

Similarly, the stability of the extremum solution is determined by the sign of the effective mass, given by  
\begin{equation}
\mu_s^2 = V_{,\rho \rho} + \epsilon H^2 R \, ,
\end{equation}
due to the absence of a turn rate in this case. Positive curvature of field space has a stabilizing effect on such solutions, while negative curvature can destabilize the extremum solution \cite{Renaux-Petel:2015mga}. 

\begin{figure}[t!]
	\centering
	\includegraphics[width=0.45\textwidth]{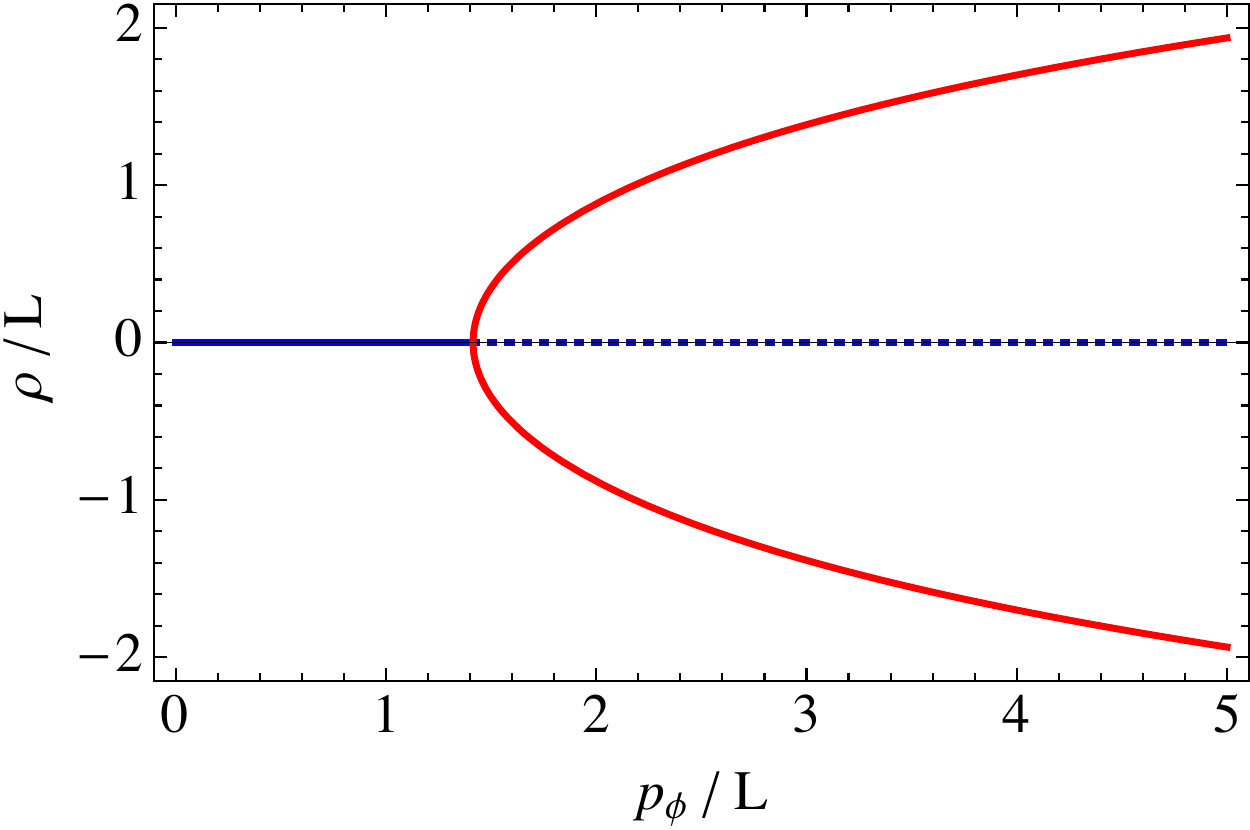}
		\includegraphics[width=0.47\textwidth]{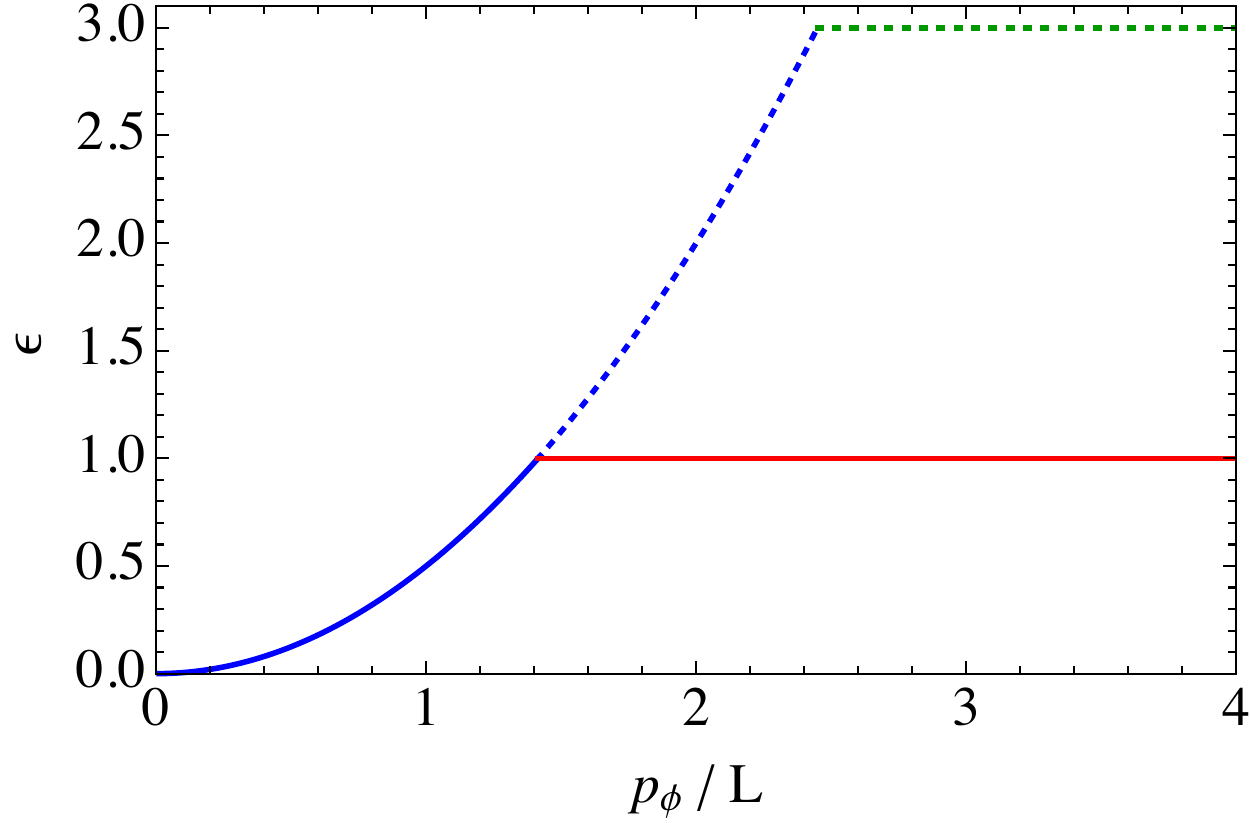}
	\caption{
{\it Left:} The bifurcation diagram for \eqref{special}. The solid (dashed) curves correspond to stable (unstable) branches. 
	{\it Right:} The value of $\epsilon$ for $\rho=0$ (blue) and $\rho\ne 0$, lying on the two branches of the pitchfork bifurcation (red). The solid (dotted) curves correspond to stable (unstable) solutions.}
	\label{fig:frozenbif}
\end{figure}

Before proceeding to connect the scaling solutions of this section to inflationary models that can be found in the literature,  we examine the transition between two solutions in a similar vein as we did in the previous section for the hyperbolic solution. In order to be specific, we take the radial dependence of the metric and potential to be 
 \begin{align} \label{special}
   f=L h = L \cosh(\rho/L) \,.
    \end{align} 
This simplifies the above discussion by having $p_\rho L_\rho =1$. This system always has an extremum solution at $\rho=0$, and in addition can also support  a frozen solution depending on the value of $p_\phi$: does $f^2 = p_\phi^2 / 2$ have roots? This leads to the critical value $p_{\rm crit} = \sqrt2 L$, with the frozen solutions only existing for supercritical gradients. The corresponding bifurcation diagram is given in Fig.~\ref{fig:frozenbif}, again providing a typical case of a supercritical pitchfork bifurcation. Depending on the initial conditions, the system will be ultimately drawn towards one of the two stable branches for $p>p_{\rm crit}$, signaling an instance of spontaneous symmetry breaking, similar to the one described in Sec.~\ref{sec:3d}.

It is worth examining the $\epsilon$ parameter for the types of pitchfork bifurcations shown in Fig.~\ref{fig:frozenbif}. 
By using Eqs.~\eqref{eq:conditionfrozen} and \eqref{eps-frozeng} we see that
 both the extremum solution, as well as the frozen one have $\epsilon=p_\phi^2 /(2f^2)$. In this work we  only consider cases where the extremum of $f$ is a minimum, which trivially leads to $\epsilon_{\rm extr}>\epsilon_{\rm froz}$. Hence, similar to the pitchfork bifurcations examined in Sec.~\ref{sec:3d}, the system in this case also ``chooses'' the solution which exhibits a smaller value of $\epsilon$ by having a non-zero turn-rate $\omega^2$.

\begin{figure}[t!]
	\centering
	\includegraphics[width=0.45\textwidth]{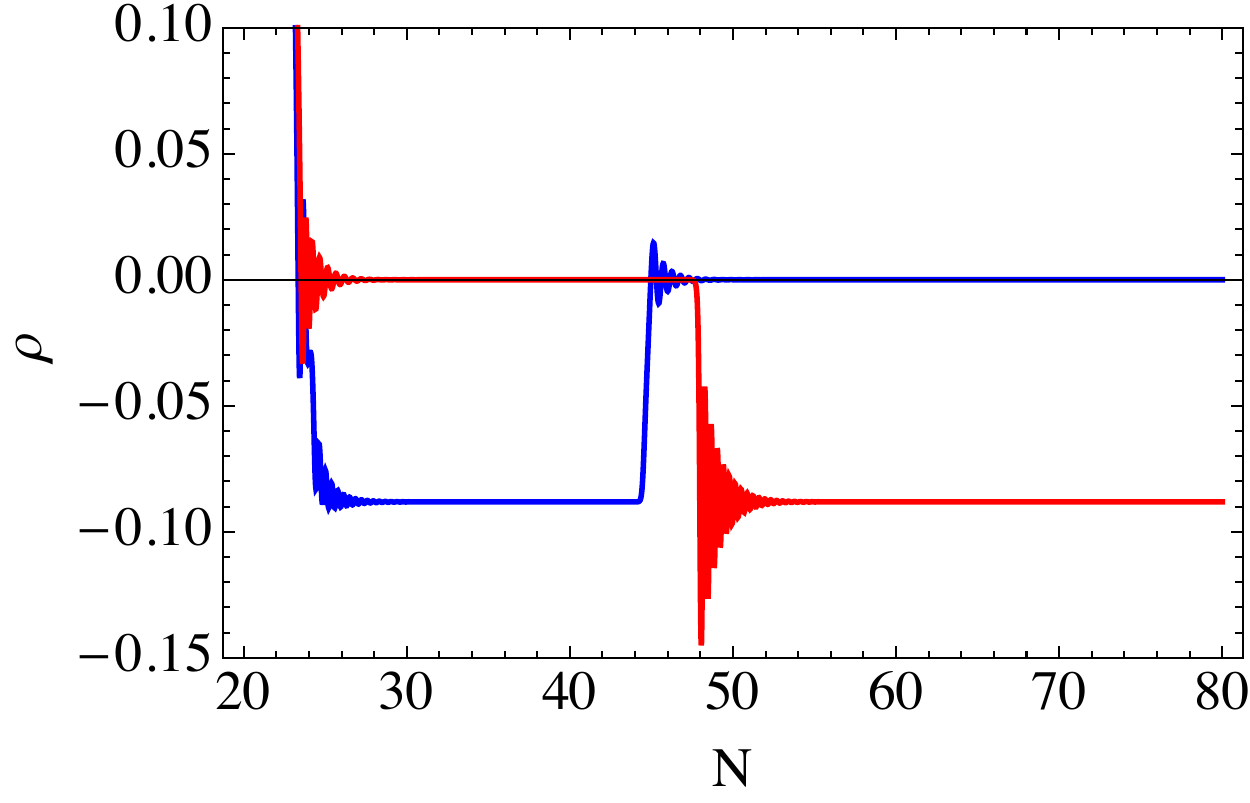}
	\includegraphics[width=0.45\textwidth]{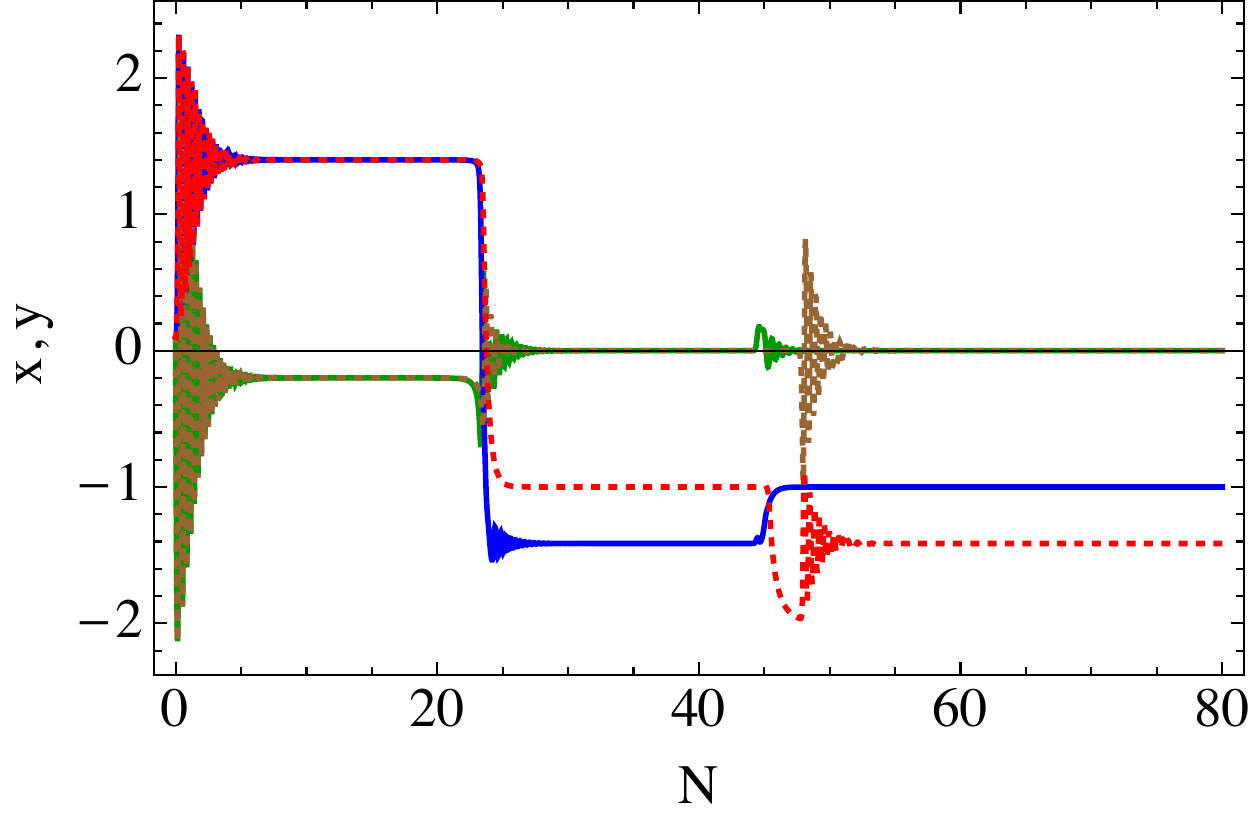}
	\caption{
	{\it Left:} Evolution of the radius $\rho$ for \eqref{special} and \eqref{eq:pphitransitionfast} with $\alpha=1$ and 
	$\{p_0,\Delta p\} = \{0.15,0.05\} ,\{0.15,-0.05\}$ 
	(red and blue respectively). The initial conditions were chosen to be far from any scaling solution.
		{\it Right:} Evolution of the normalized velocities for the same parameters. 
		Brown and red ($x$ and $y$ respectively) correspond to
		 	 $\{p_0,\Delta p\} = \{0.15,-0.05\}$
		while green and blue ($x$ and $y$ respectively) correspond to 
		$\{p_0,\Delta p \} = \{0.15,0.05\}$.}
	\label{fig:jump}
\end{figure}

As illustration we use the equivalent field-dependent gradient of Eq.~\eqref{eq:steepnessvar} ,
\beq 
p_\phi =p_0 + \Delta p\, \tanh(\alpha \phi) \,.
\label{eq:pphitransitionfast}
\eeq
arising from the ``angular'' potential
\beq
\tilde h(\phi) = e^{p_0 \phi} \left [\cosh(\alpha \phi)\right]^{\Delta p \over \alpha}
\eeq
For  $\alpha\gtrsim1$ this leads to a fast transition from sub- to supercriticality, and vice versa for $\alpha$ negative.
We initially consider $\alpha=1$ and $L=0.1$, such that $p_{\rm crit} \sim 0.14$. Fig.~\ref{fig:jump} illustrates that, for random initial conditions (field values and velocities), the system first approaches either the frozen or the extremum solution, around $N\approx 20$. When the transition described by Eq.~\eqref{eq:pphitransitionfast} occurs, around $N\approx 50$, the field positions and velocities exhibit transient oscillations, which ultimately lead to the system relaxing to the new state, controlled by the late-time value of $p_\phi$. This is reminiscent of the effects of geometric destabilization giving rise to sidetracked inflation. A minor difference is found in the fact that $\epsilon$ is constant for scaling solutions, and thus it cannot trigger geometrical destabilization, as it does in realistic inflationary set-ups \cite{Grocholski:2019mot, Garcia-Saenz:2018ifx}. Furthermore, our analysis does not take into account quantum fluctuations, which in some cases are crucial in destabilizing a system, that could classically be made to lie arbitrarily (exponentially) close to an unstable trajectory.

A further interesting feature appears for $N\lesssim 20$. In this regime, both $x$ and $y$ seem to be constant, however this behavior does not correspond to one of the scaling solutions described in this section and this phase can be considered a transient phase before the gradient / frozen solution dynamics is reached. Upon a closer examination of our set-up, our initial conditions are chosen far away from the origin, at $\rho_{\rm init}\gg 1$. This leads to the normalized angular gradient $p_\phi / f$ being exponentially small and hence the system exhibiting an approximate shift symmetry in $\phi$. This means that an initial period of hyperinflation is possible \cite{Bjorkmo:2019aev}, which is exactly the behaviour that is shown in Fig.~\ref{fig:jump} for $N\lesssim 20$.

\begin{figure}[b!]
	\centering
	\includegraphics[width=0.45\textwidth]{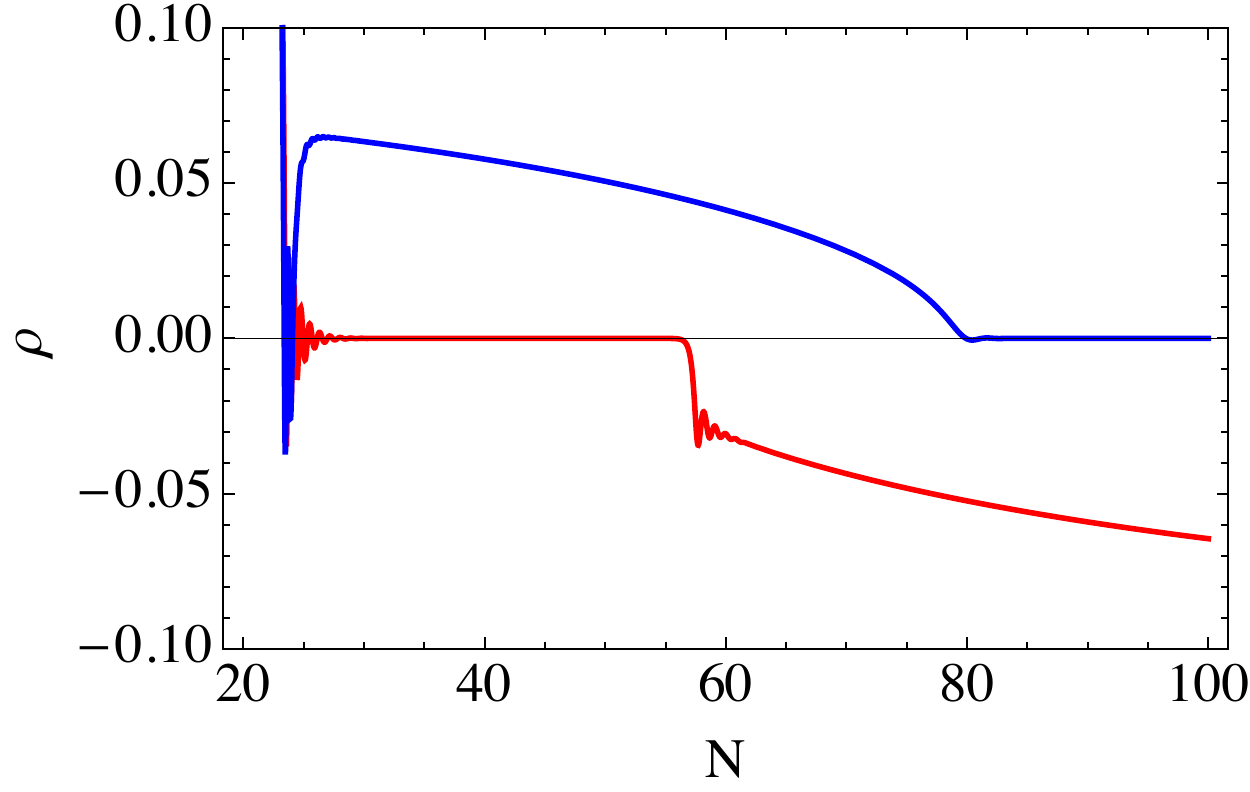}
	\includegraphics[width=0.45\textwidth]{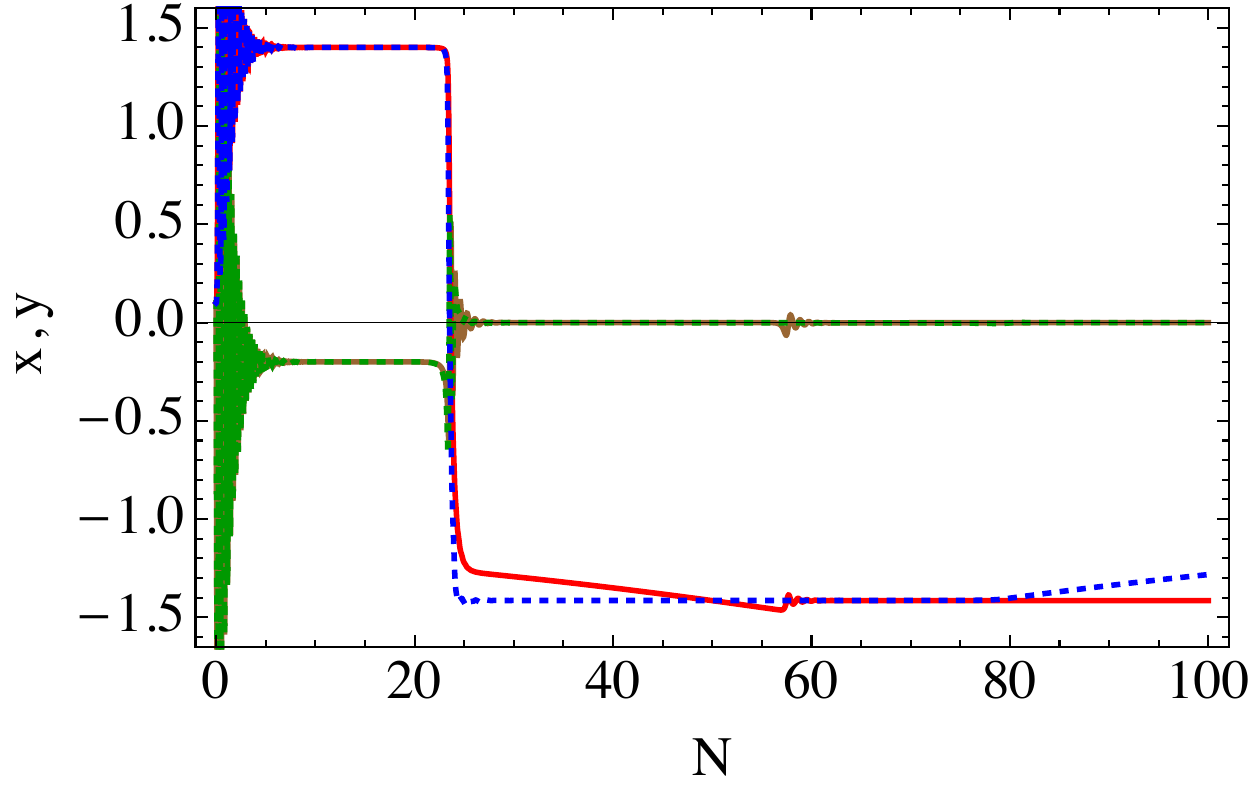}
	\caption{Identical to Fig.~\ref{fig:jump} but with $\alpha=0.01$ instead.}
	\label{fig:slowjump}
\end{figure}

While Eq.~\eqref{eq:pphitransitionfast} with $\alpha=1$ leads to a fast change in the parameters of the system and a ``jump'' between two trajectories, as shown in Fig.~\ref{fig:jump}, choosing $\alpha=0.01$ leads to a slow change in the potential steepness along the $\phi$ direction.  
Fig.~\ref{fig:slowjump} shows the resulting behavior, which is reminiscent of the behavior shown in Fig.~\ref{fig:SSB} for a radially symmetric potential with varying steepness. We clearly see that the evolution of the system adiabatically follows the change in $p_\phi$. The evolution of the coordinate $\rho$ during the frozen regime and can thus be described by the solution of Eq.~\eqref{eq:conditionfrozen} for a varying $p_\phi$.

\subsection{Comparison to inflationary models}

Realistic multi-field models of inflation are not described  by product-separable exponential potentials. However systems with general potentials and non-exponential metric functions have been shown to exhibit behaviour very similar the the ``frozen'' solution that we derived above.
 This was to be expected, as some of these models prompted the current investigation into critical points.

Non-trivial behaviour of this kind for curved geometries was first observed in multi-field $\alpha$-attractors.
It was shown in Ref.~\cite{Achucarro:2017ing} that for generic scalar potentials with moderate steepness and moderate values of the field-space curvature, the scalar field rolls towards the minimum of the potential predominantly along the radial direction $\rho$, despite the presence of possible gradients along the angular direction. Remarkably, although this is a multi-field system, the inflationary predictions come out identical as in the single-field incarnation of $\alpha$-attractors \cite{Kallosh:2013yoa}. Subsequently, it was shown that a novel phase of inflation appears at stronger curvature \cite{Christodoulidis:2018qdw}. Prompted by the angular gradient, the motion ceases to be predominantly radial at some point, and ends up in a novel dynamical attractor for a wide range of initial conditions. Moreover, it can attain a large number of e-folds in this subsequent angular phase. In this regime, the radial coordinate approximately freezes as the scalar field spins around near the boundary of the Poincare disc. The scenario of Ref.~\cite{Christodoulidis:2018qdw} can thus be seen as a succession of frozen attractors with subsequent gradients $p_\phi$ along the angular direction. The existence of a potential minimum causes inflation to end, which is not the case for scaling solutions.

It is worth reviewing the construction that led to angular inflation and compare it with the corresponding scaling solution. In Ref.~\cite{Christodoulidis:2018qdw} two scalar fields with quadratic potentials were considered on a Poincare disc. In polar coordinates with\footnote{In Ref.~\cite{Christodoulidis:2018qdw} the angular coordinate was denoted as $\theta$. Here we  call it $\phi$ to be consistent with the nomenclature used for the rest of the present analysis.}
\beq
ds^2 = {6\alpha\over (1-r^2)^2} ( dr^2 + r^2 d\phi^2) \,,
\eeq
the potential acquires a product-separable form
\beq
V(r,\phi) = \tfrac12 \alpha r^2 ( m_1^2 \cos^2\phi + m_2^2 \sin^2\phi ) \,.
\eeq
The connection to the present analysis becomes more transparent when performing the simple transformation 
\beq
r = \tanh \left (
{\rho \over \sqrt{6\alpha}}
\right ) \, ,
\eeq
which makes the radial coordinate canonically normalized
\beq
ds^2 =  dr^2 + {3\alpha\over 2} \sinh^2 \left (
\sqrt{{2 \over 3\alpha}} \rho 
\right ) d\phi^2 \,,
\eeq
which we already considered before with the identification $L^2 = 3 \alpha /2$. 
The equations of motion in the coordinates $\{\rho,\phi\}$ are\footnote{In the Poincare disc coordinates used in Ref.~\cite{Christodoulidis:2018qdw}, the radial equation of motion has an extra term of the form $\Gamma^r_{rr} \dot r^2$, which is absent when we canonically normalize the radial variable. In any case, it will not play any  role in the analysis, since $\dot r\simeq 0$ for angular inflation and $\dot r=0$ for ``frozen'' scaling solutions.}
\beqn
\ddot \phi + 3H\dot\phi  +\Gamma^\phi{\rho\phi}\dot\rho \dot\phi + {\cal G}^{\phi\phi}V_{,\phi}=0 \, , 
\label{eq:angularphieq}
\\
\ddot \rho + 3H\dot \rho  +\Gamma^\rho_{\phi \phi} (\dot\phi)^2 +V_{,\rho}=0 \, .
\label{eq:angularrhoeq}
\eeqn
The angular motion proceeds along a slow-roll trajectory derived by equating the Hubble term to the potential term in the angular equation,
\beq
\label{eq:angularSR}
3 H \dot \phi \simeq -{1\over f^2} {dV\over d\phi} \, .
\eeq
This is equal to the result given in Eq.~\eqref{frozen} by using the approximation $3 H^2 = V$. Hence in the slow-roll regime, the two constructions match for the angular motion. The radius at which angular inflation proceeds is computed by equating the Christoffel (``centrifugal'') term and the potential term in the radial equation,
\beq
\Gamma^\rho_{\phi \phi} (\dot\phi)^2 +V_{,\rho}\simeq0 \,.
\eeq
Using Eq.~\eqref{eq:angularSR} for the slow-roll motion for $\dot\phi$ along with the Christoffel term for this metric allows one to perform the following manipulation:
\beq
f^2 = {2\over 6} \left ( {V_{,\phi}\over V} \right )^2  \left ( {V \over V_{,\rho}} \right )\left ( {f_{,\rho}\over f} \right) = 
{2 p_\phi^2 \over 6 p_\rho L_\rho}  
\, ,
\eeq
where we again used the slow-roll approximation $3H^2=V$. This matches  the condition given in Eq.~\eqref{eq:conditionfrozen} for
$p_\rho L_\rho \ll 1$, which holds for strongly curved hyperbolic field-space manifolds, as the ones that were considered in Ref.~\cite{Christodoulidis:2018qdw}. 
Thus, while the trajectory of angular inflation was shown to possess a slowly varying radius, we can view it as an approximate scaling solution with slowly varying parameters for the potential steepness and a small radius of field-space curvature.

An alternative realization of the same scaling solutions arises in the sidetracked inflation scenario \cite{Garcia-Saenz:2018ifx}, employing a sum separable potential, $V = V(\rho) + V(\phi)$. After an initial phase along a geodesic trajectory, inflation is sidetracked due to the occurance of geometrical destabilization and instead ends up in a different dynamical attractor, with $\rho$ approximately constant and non-vanishing. During this phase, the background asymptotes to a frozen solution with field-dependent potential steepness $p_\phi$. It can therefore be seen as a succession of scaling attractors with slowly changing $p_\phi$. This phase has been shown to exibit an interesting EFT of  fluctuations, with either a reduced or an imaginary speed of sound $c_s$ \cite{Garcia-Saenz:2018ifx}. The distinction between the latter two possibilities depends on the subhorizon mass of isocurvature fluctuations.

Finally, a similar set-up (also based on the geometry of Eq.~\eqref{eq:isom}) was investigated very recently in Ref.~\cite{Achucarro:2019pux} with emphasis on the case of massless isocurvature flucatuations, $\mu_{\rm s}^2=0$. It turns out that there is a shift symmetry in the corresponding EFT of fluctuations and thus this was coined as ``shift-symmetric orbital inflation''. By using a product-separable potential with exponential dependence on the angle $V(\phi,\rho) = h{(\rho)} e^{p_\phi \phi}$ and demanding that scaling solutions exist for all values of the radius $\rho$ (corresponding to massless orthogonal fluctuations) and any form of the metric function $f(\rho)$, Eq.~\eqref{eq:conditionfrozen} can be integrated to give
 \begin{align}
  h(\rho) = 1 - \frac{p_\phi^2}{6 f^2(\rho)} \,.
 \end{align}
Hence by choosing the above specific relation between the metric and the potential, a scaling solution exists for any value of $\rho$. Trajectories\footnote{The exact construction of the potential relies on a multi-valued function of the angle, resulting in a ``corkscrew" structure.} 
are related by a shift symmetry $\rho \to \rho+c$ and by using this form of the potential, the stability parameter $B_{\rm froz}$ in Eq.~\eqref{eq:generalstabilityconditionfrozeng} is zero. 
This was derived in Ref.~\cite{Achucarro:2019pux} using the different perspective of the Hamilton-Jacobi formulation with angle-dependent Hubble function.
The trajectories arising in this context can thus be seen as frozen solutions with the special property of having massless isocurvature fluctuations (in the super-horizon limit). Note, that unlike the other cases with a zero eigenvalue this system leads to a non-diagonalizable Jacobian matrix, therefore stability conclusions cannot be drawn since the usual theorems do not apply.

\begin{figure}[t!]
	\centering
	\includegraphics[width=0.48\textwidth]{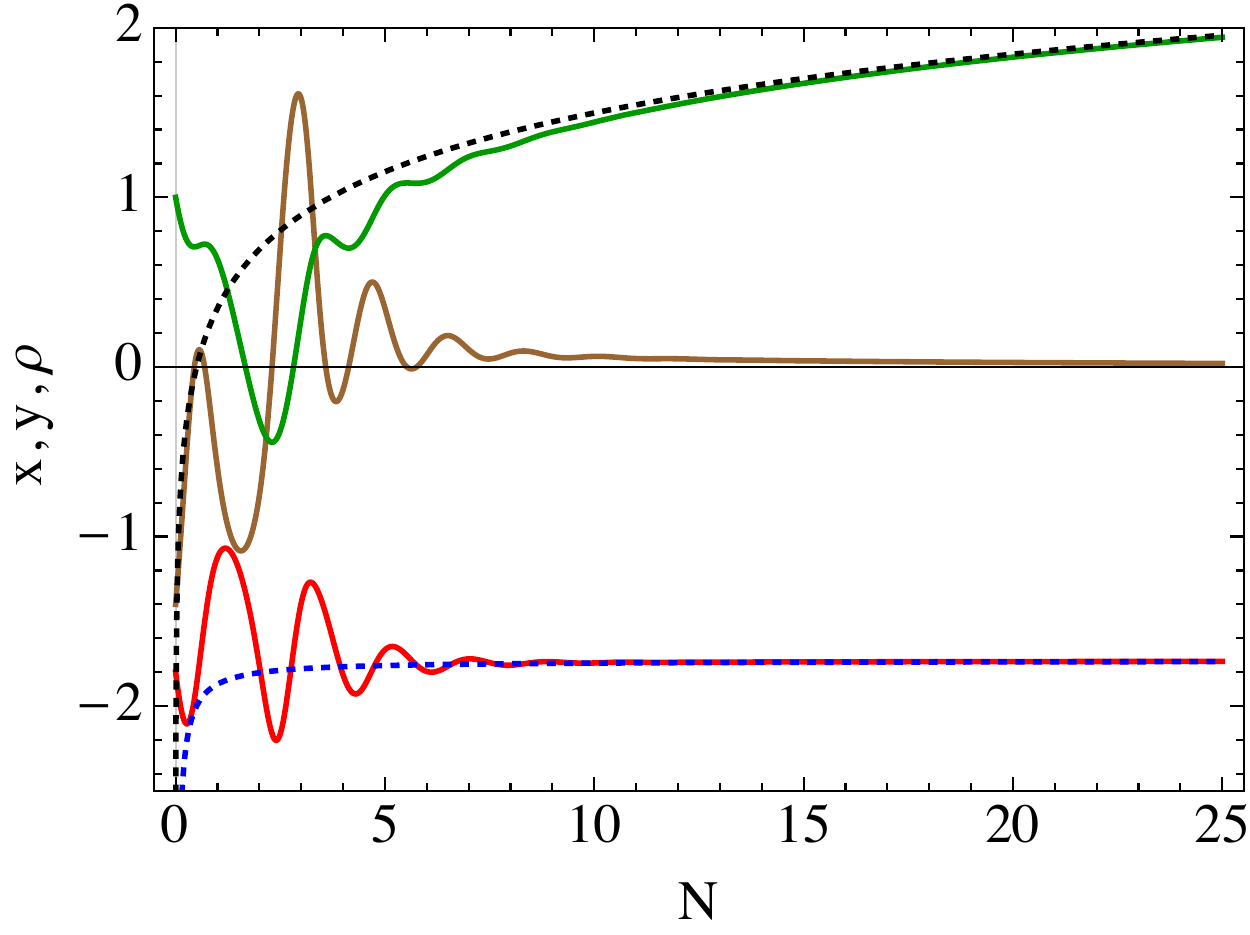}
	\includegraphics[width=0.48\textwidth]{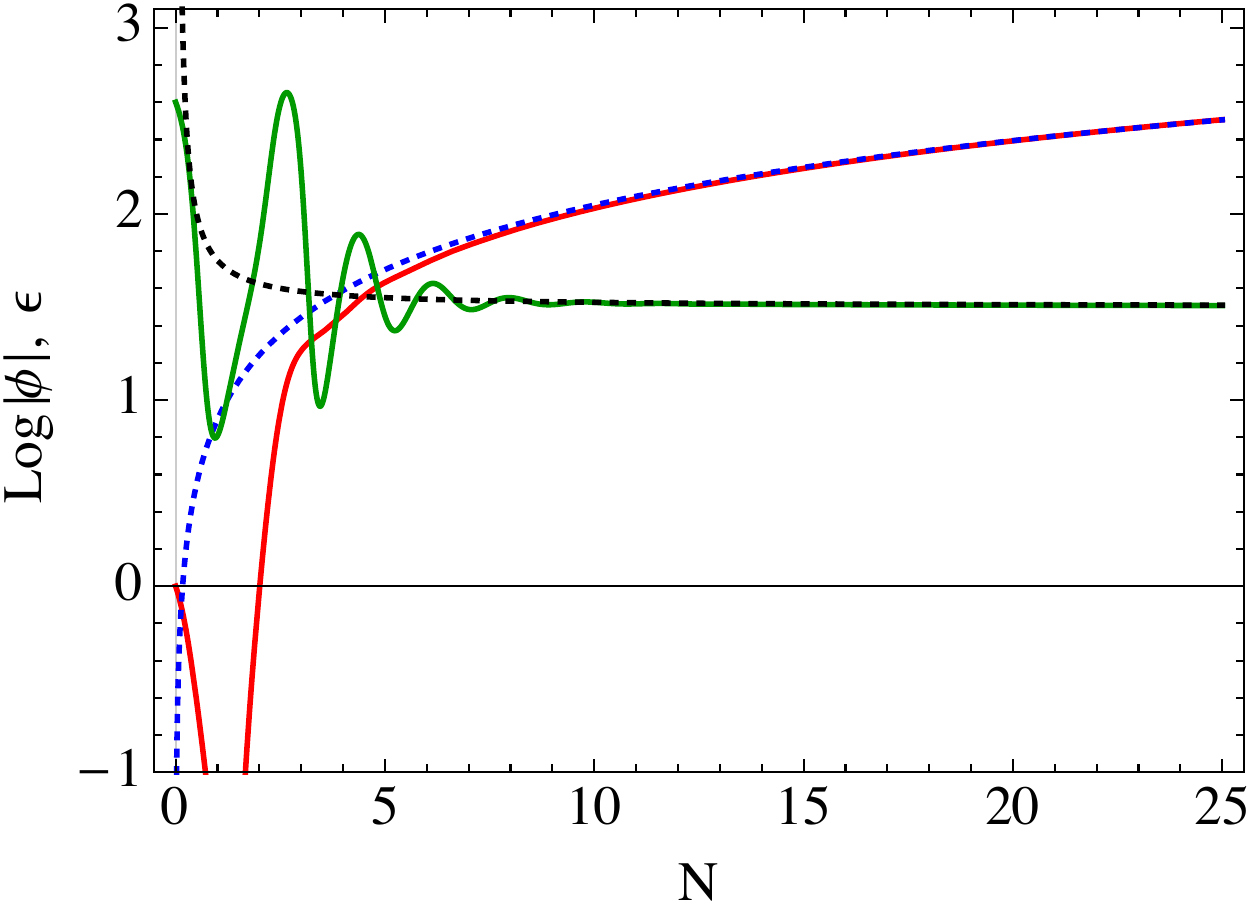}
	\caption{
	Evolution of $x,y, \rho$ ({\it left}) and $ \log|\phi|, \epsilon$ ({\it right}) for a system with an exponential metric $f = L\exp(\rho/L)$ with $L=1$ and the potential of Eq.~\eqref{eq:Vapproximatefrozeng} with $h(\rho)\propto e^{2\rho}$. The solid curves correspond to the full numerical evolution of the system and the dotted ones represent the evaluation of the scaling solution of Eqs.~\eqref{frozen} and \eqref{eq:conditionfrozen} using the field-dependent potential steepness $p_\phi \equiv V_{,\phi}/V$. Color-coding  is as follows. {\it Left:} The normalized velocities $x$ (blue solid) and $y$ (red solid, blue dotted) and the field coordinate $\rho$ (green solid, brown dotted).  {\it Right:} The logarithm of the field coordinate $\phi$ (red solid, blue dotted) and the Hubble flow parameter $\epsilon$ (green solid, black dotted). The slow evolution of the ``frozen'' coordinate $\rho$ is evident, arising from the non-exponential potential dependent of the potential on $\phi$. }
	\label{fig:frozengapprox}
\end{figure}

We conclude this section by  studying the ability of scaling solutions to describe the evolution  of a system which does not exhibit an exponential potential dependence along the $\phi$ direction and for which $h(\rho)$ does not exhibit any extrema. We choose the concrete example 
\beq
p_\phi = \sqrt{\phi^2+1} \, ,
\label{eq:approxfrozenp_phi}
\eeq
derived form the potential 
\beq
V(\phi,\rho) = V_0 \, e^{p_\rho \rho} \, e^{\frac{1}{2} \left [ \phi\sqrt{\phi ^2+1}  +\mathrm{arcsinh}(\phi )\right ]} \, .
\label{eq:Vapproximatefrozeng}
\eeq
The form of Eq.~\eqref{eq:approxfrozenp_phi} was chosen because it quickly asymptotes to a linear function $p_\phi = |\phi|$ at large field values (which makes the analytical computations easier) and is still analytic at $\phi=0$.
Fig.~\ref{fig:frozengapprox} shows the evolution of the field coordinate and derivatives for a particular realization and initial condition choice for a system described by the potential of Eq.~\eqref{eq:Vapproximatefrozeng} and an exponential metric function $f = L\exp(\rho/L)$. We see that the system undergoes a transient phase of non-scaling behavior, before settling into the attractor solution. The initial conditions were chosen so that the system is not put near the scaling solution and thus exhibits a non-trivial transient period. 
Repeating the simulation with different initial conditions resulted in a longer or shorter transient, all leading to the same attractor solution.
After the systems settles onto the approximate scaling solution, it is clearly visible that $x = d \rho / dN \approx 0$. However, the non-exponential dependence of the potential on $\phi$ leads to a slow change of $\rho$. This is captured using the expressions of the scaling solution and is reminiscent of the slowly varying radial coordinate in angular inflation \cite{Christodoulidis:2018qdw} and similar behavior in side-tracked inflation \cite{Garcia-Saenz:2018ifx}. We thus see that scaling solutions can  be an excellent starting point to study realistic inflationary scenarios, both for building intuitive understanding as well as  for constructing approximate expressions describing the full evolution.

\subsection{Hyperinflation as a special case of frozen solutions}
\label{sec:hiddenintegralsofmotion}

It is worth re-visiting our claim that the frozen solutions that are described by Eq.~\eqref{frozen} in general do not have any conserved quantities. The identical form of the Hubble flow parameter $\epsilon$ in Eqs.~\eqref{eps-frozeng} and \eqref{eq:epsilonspiralling}, provides the intriguing possibility that the two solutions are somehow related or even that one is a special case of the other, as discussed in Ref.~\cite{Christodoulidis:2019mkj}.

We start our investigation from a set-up that leads to a hyperbolic solution of Eq.~\eqref{hyper}. It only appears for a hyperbolic field-space expressed through an exponential metric function $f(\rho)$ and an exponential potential in $\rho$, while the shift symmetry along $\phi$ induces the conservation of angular momentum. Our guiding principle will be the possible reparametrisations of hyperbolic space, which can be found for example in Refs.~\cite{Carrasco:2015uma, Anguelova:2018vyr}. We start with
\beq
ds^2 = \frac{3\alpha}{2} {d\tau d\bar\tau \over (\Im \tau)^2} \, , 
\eeq 
where we choose $\tau = \phi + i e^{\rho \sqrt{2/3\alpha } }$ to describe the axion-dilaton system. This leads to the field-space line element
\beq
ds^2 = d\rho^2 + \left (L e^{\rho/L} \right)^2 d\phi^2 \, , 
\eeq
with $L = \sqrt{3\alpha/2}$. 
By choosing an alternative coordinate basis $\tau =e^Z = e^X(\cos Y + i\sin Y)$ we can easily re-write the metric as
\beq
ds^2 = L^2 {1\over \sin^2Y} \left( dX^2+dY^2 \right) \, .
\eeq
 We now canonically normalize one of the two variables, by choosing
 \beq
  \tilde \rho = L \log\left [ \tan \left ( {Y\over 2}\right ) \right ] \, , ~ X = \tilde \phi \, ,
 \eeq
  leading to
  \beq
   ds^2 = d\tilde \rho^2 + L^2 \cosh^2 \left(  { \tilde \rho \over L }\right )d\tilde \phi^2 \, ,
   \label{eq:coshmetric}
  \eeq
  which is of the form that we focus on in this paper. Both diagonal metrics define a hyperbolic space with the same Ricci curvature scalar. The inverse transformation rule is
  \beq
  \rho = -\tilde \phi L + L  \log \left [ \cosh \left( {\tilde \rho \over L}  \right) \right ] \, , ~ \phi = -e^{\tilde \phi} \tanh \left( {\tilde \rho \over L}  \right) \, .
 \eeq
 The potential that we choose in the $\{ \rho,\phi \}$ basis is one that allows for hyperbolic inflation, namely
 \beq
 V\left (\rho,\phi\right ) = V_0 e^{p \rho} \, .
 \eeq
 Following the same transformation, we  re-write the potential in the new basis $\{ \tilde \rho,\tilde\phi \}$  as
  \beq
      V\left (\tilde \rho,\tilde \phi \right ) = V_0 \left [ \cosh \left ( \tilde {\rho\over L} \right )\right ]^{p L } e^{-p L \tilde \phi}  \, .
      \label{eq:hyperINsidetrackedCoordinates}
  \eeq
  This potential is of the form of Eq.~\eqref{eq:frozenpotential} and thus allows for a frozen solution. It is straightforward to check that the time evolution of the various coordinates based on the hyperbolic and frozen solution respectively are related to each other by the above described coordinate transformations. In fact for every positive single field potential $V(\rho)$ we have
  \begin{equation}
  	p_{\tilde{\rho}} =p \tanh \left( {\tilde \rho \over L}  \right) \, ,
  \end{equation}
  and so both the potential and the metric have an extremum at $\tilde{\rho}=0$, allowing for an extremum solution.
  
We therefore showed, that while the two types of attractor solutions, hyperbolic and frozen, were discovered in different contexts and were thought to be inherently different, this is not the case, due to the enhanced symmetry properties of hyperbolic manifolds.
  In particular, hyperbolic solutions are a special case of frozen solutions, when the latter are found in hyperbolic manifolds with a specific form of the potential function $h(\tilde \rho)$ that allows for the existence of a ``hidden'' integral of motion. This allowed us to transform the metric function to a pure exponential form, while at the same time recovering an exponential potential for the dilaton field $\rho$ with a shift symmetry for the axion field $\phi$.

\begin{figure}[t!]
	\centering
	\includegraphics[width=0.48\textwidth]{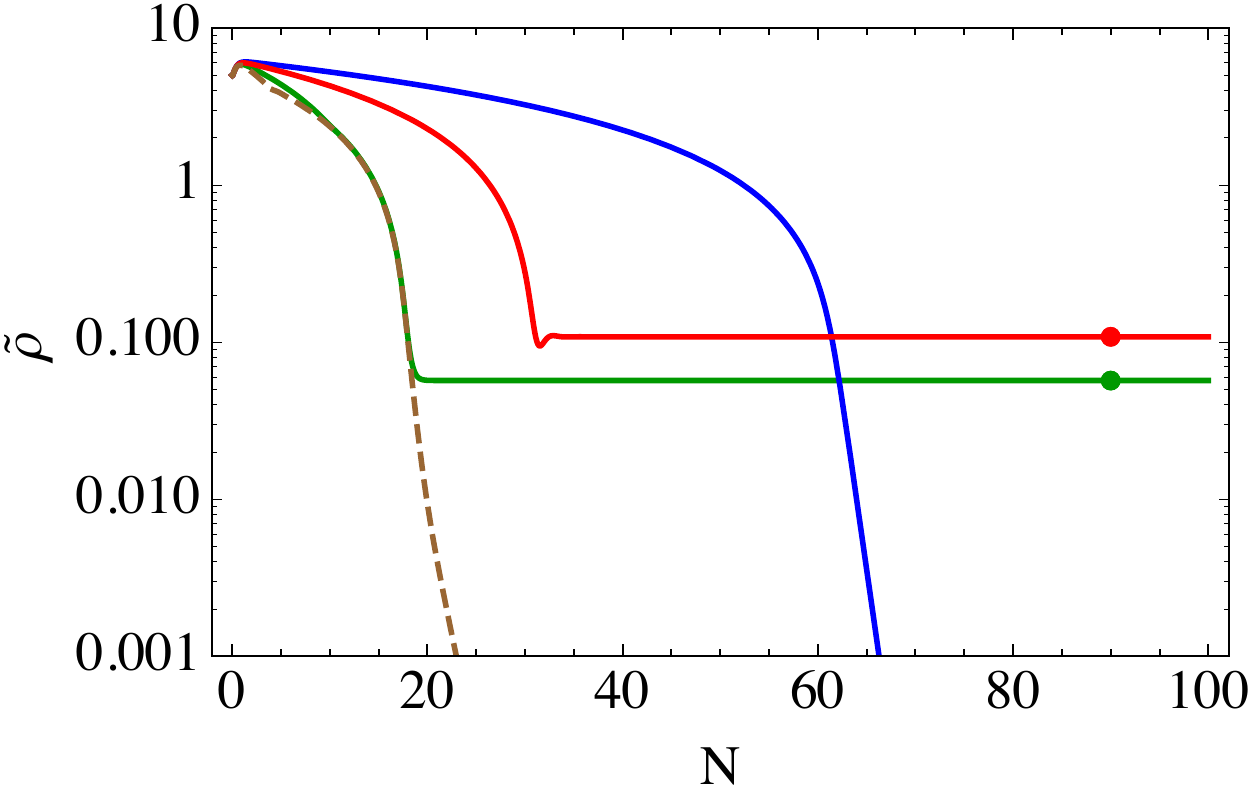}
	\includegraphics[width=0.48\textwidth]{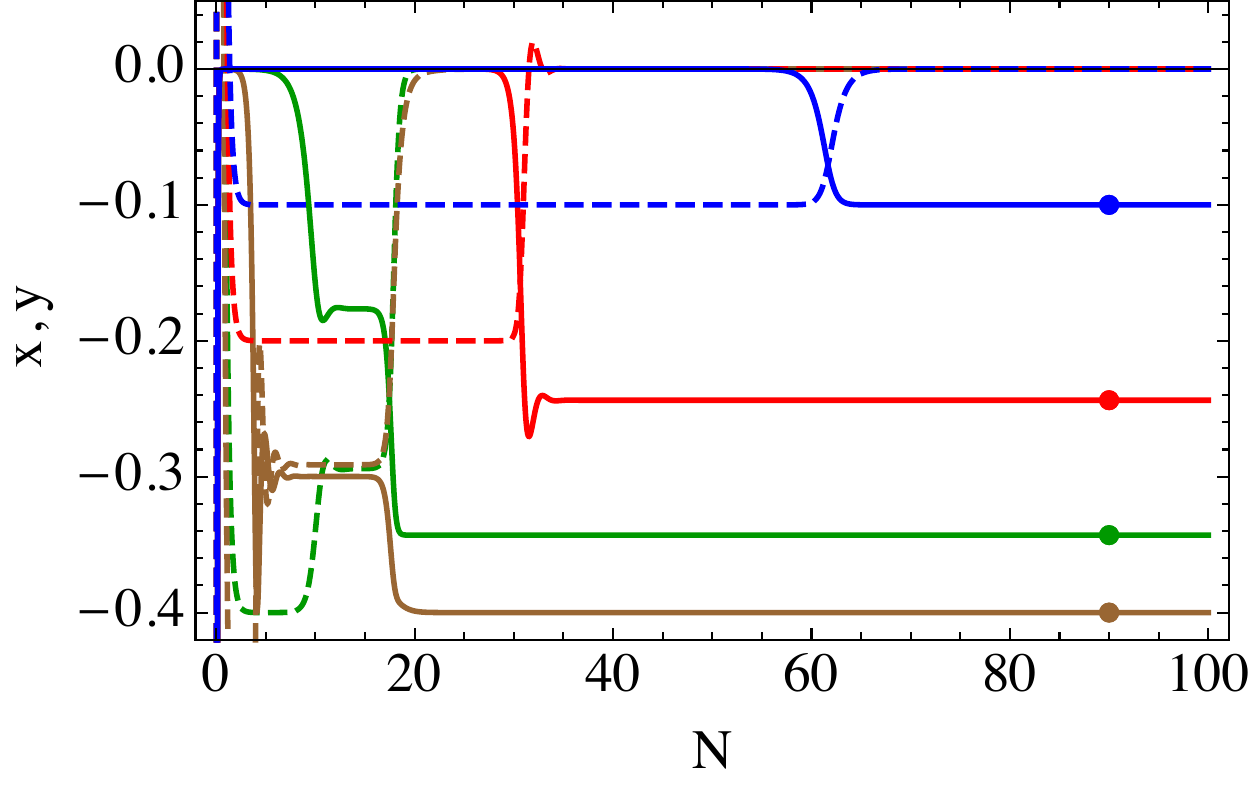}
	\caption{
	{\it Left} The evolution of the ``radial'' coordinate $\tilde \rho$ as a function of $e$-folding number $N$ for the metric of Eq.~\eqref{eq:coshmetric} and the potential of Eq.~\eqref{eq:BROKENhyperINsidetrackedCoordinates}. The parameters are $L=0.1$ and $\{p,q\}=\{ 0.1,0.1 \} $ (blue), $\{p,q\}=\{ 0.4,0.4 \} $ (red), $\{p,q\}=\{ 0.4,0.2 \} $ (green) and $\{p,q\}=\{ 0.4,0.6 \} $ (brown).  
	{\it Right:} The generalized velocity parameters $x$ (dashed) and $y$ (solid) for the same parameters and color-coding. The dots on the right part of both panels correspond to the analytical predictions for the frozen solutions. 
	 }
	\label{fig:frozengen}
\end{figure}

Before we conclude, it is worth generalizing the form of the potential given in Eq.~\eqref{eq:hyperINsidetrackedCoordinates}, such that the integral of motion is not present. A simple way is\footnote{Note that this potential can be seen to interpolate between the symmetric case of hyperinflation with $pL = qL$ and the special case with $qL=1$ discussed in the paragraph around \eqref{special}.}
    \beq
      V\left (\tilde \rho,\tilde \phi \right ) = V_0 \left [ \cosh \left ( \tilde {\rho\over L} \right )\right ]^{q L } e^{-p L \tilde \phi}  \, ,
      \label{eq:BROKENhyperINsidetrackedCoordinates}
  \eeq
where generically $q\ne p$. Taking $L=0.1$, we first point out that in the case of $q=p$ the analysis of the system in terms of $\tilde \rho$ and $\tilde \phi$ results in a hyperinflation-type solution that exists and is stable for $p\gtrsim 0.3$. We computed the evolution of the system in the basis of Eq.~\eqref{eq:coshmetric} and the results are plotted in Fig.~\ref{fig:frozengen}. We first see that for $p=q=0.1$ the system relaxes to $\tilde{\rho}=0$, while for $p=q=0.4$ the system ``freezes'' at $\tilde{\rho} \approx 0.1$, which is the exact value given by solving Eq.~\eqref{eq:conditionfrozen}. An important point is made by studying the case $p\ne q$. We see that for $p=0.4$ and $q=0.2$ or $q=0.6$, the system relaxes  to $\tilde{\rho} \simeq 0.06$ or to $\tilde{\rho}=0$ correspondingly. The former corresponds to a frozen solution, which has no exact analogue in the case of hyperinflation, since it cannot be transformed to a radially symmetric potential in the $\{\tilde \rho,\tilde\phi \}$ basis. The latter corresponds to geodesic motion along $\tilde{\rho} =0$, where both the potential and the metric exhibit a minimum.

\section{The non-isometric case: simple unified description} 
\label{sec:noniso}

The isometry assumption induces a natural distinction between the different types of solutions, and makes geometric interpretation of coordinates possible. Note that in both gradient and frozen solutions the normalized velocity of one field is fixed, while the other rolls down its exponential potential. Dropping the isometry assumption, the most general 2D metric can be written as
\begin{equation} \label{eq:nonisom}
\ud s^2 =  g^2(\rho,\phi)\ud \rho^2 + f^2(\rho,\phi) \ud \phi^2 \, ,
\end{equation}
with the following non-zero Christoffel symbols   
\begin{align}
\Gamma^{\rho}_{ \rho\rho} &=  {g_{,\rho}  \over g} \,, \qquad \Gamma^{\rho}_{ \phi \rho}=  {g_{,\phi}  \over g} \,, \qquad \Gamma^{\rho}_{ \phi \phi}= - {g g_{,\phi}  \over f^2} \,, \\   \Gamma^{\phi}_{ \phi \phi} &= {f_{,\phi}  \over f} \,, \qquad \Gamma^{\phi}_{ \rho \phi}= {f_{,\rho}  \over f}\, , \qquad  \Gamma^{\phi}_{ \rho\rho} = - {f f_{,\rho}  \over g^2} \,.
\end{align} 
The equations of motion for the two variables $x,y$, with $x=gv^{\rho}$ generalize to 
\begin{subequations}
	\label{dx4Dniso}
	\begin{eqnarray}
	x' &=& - (3-\epsilon)  \left(x + {p_{\rho} \over g} \right) - { g_{,\phi} \over f g }x y + { f_{,\rho} \over f g }  y^2   \,, \\ 
	y' &=& - (3-\epsilon)  \left( y + {p_{\phi} \over f } \right) - { f_{,\rho} \over f g }x y + { g_{,\phi} \over f g }  x^2 \,.
	\end{eqnarray}
\end{subequations}
For this problem there is no distinction between the two fields, so we will examine the existence of frozen solutions with $x=0$. In order for this to be a solution of the dynamical system we obtain the following two equations
\begin{equation}
	 (3-\epsilon) p_{\rho} =  { f_{,\rho} \over f }  y^2 \, , \qquad y = -{p_{\phi} \over f} \, .
\end{equation}
We observe the same interplay between the Christoffel and gradient terms as the isometric case, implying that the metric function $g$ does not affect the critical value for $y$. Moreover, a constant $y$ requires that $p_{\phi} / f$ is independent of $\phi$. To retain some analytical control we will consider the product-exponential potential \eqref{eq:frozenpotential} which also requires $f_{,\phi}=0$ as well \footnote{More generally one finds $V=exp[w(\rho)\int \ud \phi f] $, where $w$ is an arbitrary function.}. 

Stability is determined as usual by the eigenvalues of the Jacobian matrix. However, in order to avoid examining complicated expressions for eigenvalues we instead do the following: the characteristic equation for the Jacobian matrix is
\begin{equation} \label{eq:char}
	  \lambda \left(\lambda^3 + a_2 \lambda^2 + a_1 \lambda  + a_0 \right)= 0 \, ,
\end{equation}
where
\begin{align}
	a_2 = 2(3- \epsilon) - \sqrt{2\epsilon}{g_{,\phi} \over f g} \,, \quad
	a_1 = {1 \over 3-\epsilon}\left( (3- \epsilon) - \sqrt{2\epsilon}{g_{,\phi} \over f g} + \left({M \over H} \right)^2 \right) \,, \quad
	a_0 = \left({M \over H} \right)^2 \, ,
\end{align}
defining a mass $M$ using Eq.~\eqref{eq:generalstabilityconditionfrozeng} as $(M/H)^2 = B_{\rm froz}/4$ and which is not equal to $\mu_s^2$, when $g\neq 1$. Using Routh's criterion (see App.~\ref{app:stab}) we obtain the following two requirements that guarantee that the eigenvalues will have negative real part 
\begin{equation} \label{eq:stabcond}
	M^2>0\, , \qquad  (3- \epsilon) > - (\ln g)' =  \sqrt{2\epsilon}{g_{,\phi} \over f g}   \, . 
\end{equation} 
The latter equation defines a critical value for $\epsilon$, beyond which the solution is unstable. When $M>0$, we reproduce the stability requirements of Sec.~\ref{sec:4d}; $g=1$ corresponds to isometric frozen solutions, where also $\mu_s=M$, while $f=1$ corresponds to gradient solutions. If instead $M=0$, we factorize Eq.~\eqref{eq:char} and stability reduces to just $(3- \epsilon) + (\ln g)'>0$ and $\epsilon<3$. If $f=1$, we reproduce the stability requirements of Sec.~\ref{sec:3d}, whereas $g=1$  leads to the non-diagonalizable Hamilton-Jacobi construction.

\section{Summary and Discussion}
\label{sec:summary}

As has been illustrated in a number of investigations in recent years, slow-roll slow-turn is not necessarily the dominant behaviour in multi-field inflation, especially when the field-space manifold is curved. Depending on the curvature and the potential gradient, a range of novel regimes has been uncovered. The aim of this paper is to provide an extensive  classification of these as well as their conditions and properties.

Building on the observation that a common trait of all behaviours is a nearly constant $\epsilon$, we have outlined how {\it scaling solutions} (with constant $\epsilon$) correspond to critical points in a dynamical system and categorized various critical points on a two-dimensional scalar manifold. When the metric has an isometry these amount to:
 \begin{itemize}
 \item
  {\it Gradient solutions}, given by Eqs.~\eqref{gradient}, \eqref{grad2} and \eqref{eq:extremumsol}: these are the usual slow-roll slow-turn solutions, in which the velocity of the isometric field vanishes. In the case of a rotationally symmetric potential with a small radial gradient, the system will inflate along the direction of the gradient and hence slide down the potential landscape. This behaviour is an attractor provided the gradient is subcritical \eqref{critical}, and does not require an isometry direction (see Sec.~\ref{sec:noniso}). Similarly, provided a non-rotationally symmetric potential has a minimum along either its radius-like or its angle-like direction, one can see the same slow-roll slow turn behaviour for the orthogonal field.
  Models with an isometry will 
 evolve close to a gradient solution if $p_{\rho}' \ll 1$ and $p_{\phi} \approx 0$.
  
\item
{\it Hyperbolic solutions}, given by Eq.~\eqref{hyper}: if the radial gradient exceeds a certain threshold and if the scalar geometry is hyperbolic, the solution enters a spiralling phase, with a non-vanishing velocity along both the radial and angular direction in such a way as to preserve angular momentum. This phase therefore exhibits spontaneous symmetry breaking and is an attractor provided the gradient is supercritical, as given by Eq.~\eqref{critical}. Under a coordinate transformation this can be mapped to a frozen solution (see below).

\item
{\it Frozen solutions}, given by Eqs.~\eqref{eq:conditionfrozen} and \eqref{frozen}: generic potential gradients allow for a frozen solution with only angular-like velocity, while the radial coordinate is ``frozen'' at a non-vanishing value. Stability of this critical point can be translated into a positive mass of its superhorizon isocurvature fluctuations, and this solution does not require an isometric geometry (see Sec.~\ref{sec:noniso}).  A model with an isometry will evolve close to a frozen solution if $p_{\phi}' \ll 1$.
 
\item 
{\it Kinetic solutions:} finally,  solutions that are dominated by the kinetic energy have $\epsilon = 3$ and are possible for all geometries and gradients. A kinetic solution will only be an attractor whenever the frozen (or hyperbolic) solutions do not exist and the radial gradient remains subcritical with $p_{\rho}\geq \sqrt{6}$.
\end{itemize}
Furthermore, as we have outlined, the hyperbolic solution can be mapped by a coordinate transformation to the frozen form and can be seen as a special case (with additional symmetries and conservation of angular moment) of this class. Consequently, all scaling behaviour is either of the slow-roll slow-turn type along a gradient and a geodesic, or can be formulated as slow-roll along one direction while the orthogonal direction is frozen, or stationary. The same result was argued for general inflationary scenarios in Ref.~\cite{Christodoulidis:2019mkj}.

The scaling attractors can be seen as the generalization of De Sitter solutions with a non-vanishing Hubble flow parameter $\epsilon$. Remarkably, it is possible to link the known non-slow-roll slow-turn inflationary models in these categories: even when these models do not have the exponential potential required for an exact scaling solution, they can often be approximated as an exponential with slowly varying steepness. To a very good accuracy, the resulting trajectory  can then be approximated by a scaling solution with field-dependent parameters. It is in this sense that hyperinflation provides an approximation of a hyperbolic solution, and angular, sidetracked and orbital inflation asymptote to frozen solutions, at least for part of their evolution. 

The formulation in terms of (critical points of) a dynamical system allows for a very simple organizing principle for these novel behaviours. Their stability is easily calculated from a linearised analysis around these critical points, leading to conditions \eqref{eq:stabcond} which do not necessarily include the effective mass on super-Hubble scales $\mu^2_s$. The condition $\mu_s^2>0$ becomes necessary and sufficient for background stability when there exists a coordinate system in which the orthogonal field is canonically normalized. In many cases, stereotypical pitchfork bifurcations take place, e.g.~as a function of the radial gradient in a hyperbolic geometry. Remarkably, in all pitchfork bifurcations that we encountered, the instability will always drive the system to the critical point with the smallest possible value of $\epsilon$.   Moreover, while we have performed a classification of scaling solutions for two-dimensional geometries it would be interesting to investigate whether this behaviour is also present in higher-dimensional geometries with less isometries than the hyperbolic one.

Finally, as cosmological obervations involve fluctuation correlations, it is essential to go beyond the background description and perform an analysis of perturbations. As we have indicated in the previous sections, this has been performed in some of the models that exhibit  attractors that violate the slow-roll slow-turn conditions, with interesting properties at the EFT level. It would be worthwhile to investigate what the most convenient description is for exact scaling solutions, and what this adds to the fluctuation analyses. This is  left for future work.

\section*{Acknowledgements}

It is a pleasure to thank S\'{e}bastien Renaux-Petel and Sebasti\'{a}n Garc\'{i}a-S\'{a}enz for helpful discussions. The authors gratefully acknowledge support from the Dutch Organisation for Scientific Research (NWO). PC would also like to thank the Institut d'Astrophysique de Paris for its warm and generous hospitality.

\appendix
\section{Coordinate transformations and isometries} \label{app:isom}

A metric with an isometry possesses a Killing vector $\vec{\xi} = \xi^I \partial_I$. Under a coordinate change $x^I\equiv \{ \chi,\psi \} \rightarrow \tilde{x}^I \equiv \{ \rho,\phi \} $ the components of the vector transform according to 
\begin{equation}
	\tilde{\xi}^I = {\partial \tilde{x}^I \over \partial x^K}\xi^K \, .
\end{equation}
Therefore, it is possible to set the first component to zero which is equivalent to solving the advection equation with variable coefficients
\begin{equation}
	{\partial \rho \over \partial \chi}\xi^{\chi} + {\partial \rho \over \partial \psi}\xi^{\psi} = 0 \, .
\end{equation}
With appropriate boundary conditions this can  always be solved, e.g. by the method of characteristics. Thus, in the new coordinate system the Killing vector points along the second basis vector $\vec{\xi} = \tilde{\xi}^{\phi} \partial_{\phi}$ and so the new metric is independent of that coordinate. Any remaining off-diagonal terms can be absorbed through a redefinition of the  variable $\phi$, whereas ${\cal G}_{\rho \rho}$ can be set to one by an appropriate redefinition of $\rho$
\begin{equation}
\tilde{\phi} = -\int {{\cal G}_{\rho\phi} \over {\cal G}_{\phi\phi}} \ud \rho + c_1(\phi) \qquad \tilde{\rho} = \int \sqrt{{\cal G}_{\rho\rho}}  \ud \rho + c_2(\phi) \, .
\end{equation} 
So, indeed, the most general form of the metric with one isometry is of the form of Eq.~\eqref{eq:isom}. As a side note, 2D metrics can have 0, 1 or 3 isometries, where the latter describes a maximally symmetric space of constant curvature (flat, hyperbolic or spherical).

\section{Hurwitz-Routh stability criterion} \label{app:stab}
A polynomial of degree n is called stable if all roots have negative real part. The relevance to the stability of dynamical systems is clear: the characteristic equation of the $N\times N$ Jacobian matrix is an n\textsuperscript{th} order polynomial 
\begin{equation}\label{eq:nth}
\lambda^n + a_{n-1} \lambda^{n-1} + \cdots + a_0= 0 \, ,
\end{equation}
and if every root has negative real part the dynamical system is called (asymptotically) stable. Analytical formulae for the roots of Eq.~\eqref{eq:nth} exist up to 4\textsuperscript{th} order so it is necessary to develop tools to infer stability without finding the roots. One method is the Hurwitz-Routh theorem \cite{2008arXiv0802.1805B}: a polynomial will be stable if every coefficient is positive $a_n>0$ and if every principal Hurwitz determinant is also positive. The latter is the determinant of a matrix constructed as follows: the first elements are $\{ a_{n-1},1 \}$ while the rest are zeros. In the second row the first elements are $\{a_{n-3},a_{n-2},a_{n-3},1 \} $ and the rest zero. Similarly, the $i$-th row is constructed using the $a_{n-i}$ coefficient 
\begin{equation}
\Delta_k = \begin{pmatrix}
a_{n-1} &1 &\cdots	&0 \\
a_{n-3} &a_{n-2} &\cdots	&0  \\
\cdots	&\cdots &\cdots &0  \\
a_{n-k} &\cdots	&a_{k+1}  &a_{k}
\end{pmatrix} \, .
\end{equation}
The criterion is formulated as follows: $|\Delta_k|>0$, for all $k<n$.

For a quadratic equation the criterion reduces to positivity of every coefficient, while for a cubic equation we obtain the additional condition $a_{2} a_{1}> a_{0}$.

\section{Bifurcations in dynamical systems }
When a dynamical system depends continuously on some parameters then stability of critical points may depend on the parameter values. More specifically, variation of the parameters may alter stability properties of certain critical points or it can lead to the creation/annihilation of critical points with different stability properties. 

Specializing to one dimension
\begin{equation}
	\dot{x} = f(x,a) \, ,
\end{equation}
a necessary (but not sufficient) condition for the existence of a bifurcation at a critical point located at $x=0$ and for the bifurcation parameter $a_{crit}=0$ is \cite{strog,bif}
\begin{equation}
	f(0,0) = 0 , \qquad {\partial f \over \partial x}(0,0) =0\, . 
\end{equation} 
The next-to-leading terms in the Taylor expansion near the critical point will determine the type of bifurcation \footnote{With a redefinition of $x$ the constant $c_2$ can be set to $\pm 1$, while $c_1$ can be absorbed in the definition of the bifurcation parameter.}:
\begin{itemize}
	\item if $\partial_{xx}f(0,0)\neq 0$ and $\partial_{xa}f(0,0)\neq 0$ then a \textit{transcritical bifurcation} occurs for $a=0$. The normal form of equations around the critical point is
	\begin{equation}
		\dot{x} = c_1 a x + c_2 x^2 \, .
	\end{equation}
	
	\item if $\partial_{xx}f(0,0) = 0$ instead but $\partial_{xa}f(0,0)\neq 0$ and $\partial_{xxx}f(0,0)\neq 0$ then a \textit{pitchfork bifurcation} occurs for $a=0$. Similar the normal form is
	\begin{equation}
	\dot{x} = c_1 a x + c_2 x^3 \, .
	\end{equation}
	
\end{itemize}
An example of the first kind is the exchange of stability between the solution $v=-p$ and the kinetic solution for one field when $p=\sqrt{6}$. Translating the critical point at the origin by defining $z=v+p$ and changing the bifurcation parameter to $k=p-\sqrt{6}$ Eq.~\eqref{eq:2Dsystem2} becomes
\begin{equation}
 z'= \sqrt{6} k z + \sqrt{6}z^2 + \left( {1 \over 2} k^2 z + kz^2 \right) \, .
\end{equation}
Note that there are no first order terms in either $k$ or $z$, the second order terms are exactly those mentioned above and the terms in parenthesis are higher order.

An example of the second kind is the bifurcation of the velocity in the 3D hyperbolic problem. After a similar  coordinate translation the 2D reduced dynamical system can be transformed to
\begin{align} \label{dzcm}
	z' &=  \frac{y^2}{L} -\frac{1}{L}\left( \sqrt{\frac{1}{L^2}+6} - \frac{1}{L}\right) z + \left( \sqrt{\frac{1}{L^2}+6} - \frac{1}{L} \right) k z + \frac{1}{2} \left( k^2 + y^2\right)z  \\ \nonumber & - \left( \sqrt{\frac{1}{L^2}+6 } - \frac{1}{L} + k \right) z^2  + \frac{z^3}{2}  \, , \\ \label{dycm}
	y' &=  y \sqrt{\frac{1}{L^2}+6}  \left( k - z \right) + \frac{k^2 y}{2} - k y z -\frac{y^3}{2} \, , \\ 
	k'&=0 \, ,
\end{align}
where the last equation increases the dimension of the system in order to study the bifurcation with center manifold techniques. Parametrizing the stable direction in terms of the center manifold variables $z=z(y,k)$ we obtain
\begin{equation} \label{dzcm2}
 z' = y' {\partial z \over \partial y} \, ,
\end{equation}
with ${\partial z \over \partial y}(0,0)={\partial z \over \partial k}(0,0)=0$. Thus, $z$ can be expressed as a Taylor expansion over the two variables as
\begin{equation}
	z = c_{11}y^2 + c_{12} yk+ c_{22}k^2 + \cdots \, .
\end{equation}
Since $z$ is at least second order in terms of $y$ \footnote{Close to the critical point $y\sim \sqrt{k}$, so $k$ is of order $y^2$.} center manifold dynamics to lowest order is governed by 
\begin{equation}
	y' =  \sqrt{\frac{1}{L^2}+6}  y k  -\frac{y^3}{2} + O(y^4)\, ,
\end{equation}
and a pitchfork bifurcation happens at $(y,k)=(0,0)$.

For the 4D problem the frozen solution a bifurcation is possible if $f_{,\rho}$ and $p_{\rho}$ have a common root which without loss of generality we consider to be at $\rho=0$. Diagonalizing the system the zero eigenvalue occurs for the variable 
\begin{equation}
	z = \rho + {x \over 3-\epsilon_c} \, .
\end{equation}
while the other two variables are $x/(3-\epsilon_c)$ and $w= y - p_{\phi}/f(0)$. The equation of motion for $z$ is given by
\begin{equation}\label{dz_cm_pre}
z'  = x\left( 1 - {3-\epsilon \over 3-\epsilon_c} \right) - V_{eff}^{,\rho}\, ,
\end{equation}
where $V_{eff}^{,\rho}$ is the effective potential introduced in \cite{Tolley:2009fg} and extensively studied in \cite{Christodoulidis:2019mkj}. As usual we consider $x$ and $w$ to be quadratic functions of $z$ and the bifurcation parameters and the first term of Eq~\eqref{dz_cm_pre} is at least 4\textsuperscript{th} order. Therefore, close to the critical point $V_{eff}^{,\rho}$ determines the dynamics of the center manifold. Expanding around $z=0$ we obtain to lowest order
\begin{align}\label{dz4d}
	z ' &= - {k + \sqrt{2 \epsilon_c} (p'(0) - R) w \over 3 - \epsilon_c} z + {1 \over (3 - \epsilon_c)}\left(-{R^2 \epsilon_c \over 4}  - {1 \over 6} (3 - \epsilon_c) p^{(3)}(0) + {1 \over 3} {f^{(4)} \over f} \epsilon_c \right)z^3 \, ,\\ \label{dy4d}
	w'&= - (3 - \epsilon_c) \left( w + {\sqrt{2 \epsilon_c} \over 4}R \right)z^2 \, ,
\end{align}
where we omitted the equation for $x$ as it does not affect the equation of the center manifold.
 Since $z'$ is at least 3\textsuperscript{rd} order in $z$ the quadratic coefficient of $w(z)$ should cancel the second term of Eq.~\eqref{dy4d} and so $w(z)= - {1 \over 4}\sqrt{2 \epsilon_c} R$. Substituting back to Eq.~\eqref{dz4d} we finally obtain the equation of the center manifold
\begin{equation}
(3 - \epsilon_c) z ' = - k z + \left({2 p' -3R^2 \epsilon_c \over 4}  - {1 \over 6} (3 - \epsilon_c) p^{(3)} + {1 \over 3} {f^{(4)} \over f} \epsilon_c \right) \Big|_{\rho=0} z^3 \, ,
\end{equation}
and a supercritical pitchfork bifurcation occurs for $k=0$. This equation is exactly the same as the expansion of $V_{eff}^{,\chi}$ around $\rho=0$ up to 3\textsuperscript{rd} order when we have expressed every variable in terms of $\rho$.

\end{document}